\documentclass[twocolumn,prd,aps,showpacs,preprintnumbers,nofootinbib]{revtex4}

\usepackage{amsmath,amsfonts,amssymb}
\usepackage{bm}		       
\usepackage{wrapfig}           
\usepackage{graphicx}          
\usepackage{color}             
\usepackage{dcolumn}           
\usepackage{psfrag}
\usepackage{pstricks,pst-plot} 

\usepackage{t1enc}
\usepackage[english]{babel}   
\usepackage[latin1]{inputenc}

\allowdisplaybreaks

\begin{document}

\title{Phasing of gravitational waves 
from inspiralling eccentric binaries
at the third-and-a-half post-Newtonian order}

\author{Christian K\"onigsd\"orffer}
\email{C.Koenigsdoerffer@uni-jena.de}
\affiliation{Theoretisch-Physikalisches Institut,
Friedrich-Schiller-Universit\"at Jena, 
Max-Wien-Platz 1,
07743 Jena, Germany}

\author{Achamveedu Gopakumar}
\email{A.Gopakumar@uni-jena.de}
\affiliation{Theoretisch-Physikalisches Institut,
Friedrich-Schiller-Universit\"at Jena, 
Max-Wien-Platz 1,
07743 Jena, Germany}

\date{\today}

\begin{abstract}
We obtain an efficient description for the dynamics of
nonspinning compact binaries moving in inspiralling eccentric orbits
to implement the phasing of gravitational waves
from such binaries at the 3.5 post-Newtonian (PN) order.
Our computation heavily depends on the phasing formalism,
presented in 
[T.~Damour, A.~Gopakumar, and B.~R.~Iyer,
Phys. Rev. D \textbf{70}, 064028 (2004)],
and the 3PN accurate generalized quasi-Keplerian parametric solution
to the conservative dynamics of nonspinning compact binaries
moving in eccentric orbits, 
available in 
[R.-M.~Memmesheimer, A.~Gopakumar, and G.~Sch\"afer,
Phys. Rev. D \textbf{70}, 104011 (2004)].
The gravitational-wave (GW) polarizations $h_{+}$ and $h_{\times}$
with 3.5PN accurate phasing should be useful
for the earth-based GW interferometers, 
current and advanced,
if they plan to search for gravitational waves
from inspiralling eccentric binaries.
Our results will be required to do \emph{astrophysics}
with the proposed space-based GW interferometers
like LISA, BBO, and DECIGO.
\end{abstract}


\maketitle

\section{Introduction}

\label{phasing:IntroSec}

Inspiralling compact binaries of arbitrary mass ratio
moving in \emph{quasi-circular} orbits are the most plausible
sources of gravitational radiation for the first generation
ground-based interferometric detectors \cite{GWIFs}.
The availability of highly accurate general relativistic 
theoretical waveforms required to extract the 
weak GW signals from the noise-dominated 
interferometric data is the main reason 
for the above understanding. 
The dynamics of long lived and isolated compact binaries
can be modelled accurately in the 
PN approximation 
to general relativity as point particles 
moving in quasi-circular orbits.
The PN approximation allows one to express the equations of motion
of a compact binary as corrections to the Newtonian equations of motion
in powers of $(v/c)^2 \sim G M / (c^2 R)$,
where $v$, $M$, and $R$ are
the characteristic orbital velocity,
the total mass, and the typical orbital separation of the binary,
respectively.
Recently, the orbital evolution of nonspinning compact binaries
in quasi-circular orbits,
under the action of general relativity,
was computed up to the 3.5PN order in Ref.~\cite{ref1NEW}.
The amplitude corrections to the GW polarizations
$h_{+}$ and $h_{\times}$ are also available 
to the 2.5PN order \cite{ABIQ04}.

However, a recent surge    
in astrophysically motivated investigations
indicates that compact binaries of arbitrary mass ratio
moving in inspiralling \emph{eccentric} orbits are
also plausible sources of gravitational radiation \emph{even}
for the ground-based GW interferometers.
One of the earliest scenarios involves Kozai oscillations,
associated with hierarchical triplets
that may be present in globular clusters \cite{K62,MH02,Rasio2000,W03}.
Last year, it was pointed out 
that during the late stages of 
black hole--neutron star \mbox{(BH--NS)} inspiral
the binary can become eccentric \cite{DLK05}.
This is because in general 
the neutron star is not disrupted at the first phase of mass transfer
and what remains of the neutron star is left on a wider eccentric orbit
from where it again inspirals
back to the black hole.
This scenario was very recently invoked to explain the       
light curve of 
the short gamma-ray burst GRB $050911$ \cite{Page05}.
Another scenario,
reported in \emph{Nature},
suggests that at least partly
short GRBs are produced by the merger of \mbox{NS--NS} binaries,
formed in globular clusters 
by exchange interactions 
involving compact objects \cite{GZM_nature_2006}.
A distinct feature of such binaries
is that they have high eccentricities 
at short orbital separation 
[see Fig.~2 in Ref.~\cite{GZM_nature_2006}].
Compact binaries that merge 
with some residual eccentricities may
be present in galaxies too.
Chaurasia and Bailes
demonstrated that a natural consequence of an asymmetric kick 
imparted to neutron stars at birth 
is that the majority of \mbox{NS--NS} binaries 
should possess highly eccentric orbits \cite{CB05}.
Further, the observed deficit of highly eccentric short-period 
binary pulsars was attributed to selection effects in pulsar surveys.
The authors also pointed out that their conclusions are
applicable to \mbox{BH--NS} and \mbox{BH--BH} binaries. 
Yet another scenario
that can create inspiralling eccentric binaries
with short periods
involves compact star clusters.
It was noted that the interplay
between GW-induced dissipation and stellar scattering
in the presence of an intermediate-mass black hole
can create short-period highly eccentric binaries \cite{HA05_ApJ}.
Finally, 
a very recent attempt
to model realistically compact clusters
that are likely to be present in galactic centers
indicates that compact binaries 
usually merge with eccentricities \cite{RS_MNRAS}.
These above mentioned scenarios    
force us to claim that compact binaries 
in inspiralling eccentric orbits
are plausible sources of gravitatinal waves
\emph{even} for the ground-based GW interferometers.

In order to do \emph{astrophysics}
with the proposed space-based GW interferometers,
LISA \cite{LISA}, 
BBO \cite{BBOproposal}, and 
DECIGO \cite{DECIGO},
it is required to have highly accurate 
GW polarizations, $h_{+}$ and $h_{\times}$,
from compact binaries of arbitrary mass ratio
moving in inspiralling eccentric orbits.
Recall, the earlier discussions also indicate 
that stellar-mass compact binaries
in eccentric orbits
are excellent sources for LISA.
Furthermore, 
it is expected that LISA will
``hear'' gravitational waves from 
intermediate-mass black holes
moving in 
highly eccentric orbits \cite{Gultekin05,M05,GFR05}.
Finally, several papers which appeared recently in the \emph{arXiv}
indicate that supermassive black-hole binaries,
formed from galactic mergers, may coalesce with 
orbital eccentricity \cite{SA02,RS_SMBHbinary,BLS02,AN05,IFM05}.
It is interesting to note that these investigations
employ different techniques and astrophysical scenarios
to reach the above conlusion.

The above mentioned astrophysically
inspired investigations motivated us to extend the phasing formalism,
developed and implemented with 2.5PN accuracy in Ref.~\cite{DGI},
to the next PN order, namely, the 3.5PN order.
The phasing formalism provides
a method to construct, almost analytically, 
templates for compact binaries of arbitrary mass ratio
moving in inspiralling eccentric orbits.

We recall that accurate templates
for the detection of gravitational waves require \emph{phasing},
i.e., an accurate mathematical modelling of the 
continous time evolution of the GW polarizations. 
In the case of inspiralling eccentric binaries,
the above modelling requires
the combination of three different time scales
present in the dynamics, namely,
those associated with
the radial motion (orbital period),
advance of periastron, and 
radiation reaction,
without treating radiation reaction in an adiabatic mannner.
In Ref.~\cite{DGI}, 
an improved \emph{method of variation of constants}
was presented to combine these three time scales
and to obtain the phasing at the 2.5PN order.
We note that the techniques adapted in Ref.~\cite{DGI} 
were influenced by the mathematical formulation,
developed by Damour \cite{TD82,TD83,TD_F85},
that gave the (heavily employed)
accurate relativistic \emph{timing formula}
for binary pulsars \cite{DD86,DT92}.

It is possible to extend the phasing to the 3.5PN order,
mainly because of the recent
determination of the 3PN accurate
generalized quasi-Keplerian parametric solution 
to the conservative dynamics
of nonspinning compact binaries of arbitrary mass ratio
moving in eccentric orbits \cite{MGS}.
This parametrization,
presented in Ref.~\cite{MGS},
allows one to solve analytically the 3PN accurate 
conservative dynamics of nonspinning compact binaries,
computed both in ADM-type coordinates \cite{DJS00A} 
and in harmonic coordinates \cite{BF01},
indicating the deterministic nature 
of the underlying dynamics \cite{GK05nochaos}.
Further,
we recall that Ref.~\cite{MGS}
extends the quasi-Keplerian parametrization
developed by Damour and his collaborators \cite{DD85,DS88,SW93},
which is crucial to construct the timing formula
relevant for relativistic binary pulsars \cite{DD86,DT92}.
This observation clearly reveals the link, 
as envisaged by Damour,
connecting GW observations
of inspiralling compact binaries 
to the timing of binary pulsars.
In Ref.~\cite{DGI}, explicit computations
to realize the phasing at the 2.5PN order
were done in ADM coordinates as at that time 
the 2PN accurate generalized quasi-Keplerian parametrization,
required by the method of variation of constants,
was only available in ADM coordinates.
However, in this paper,
computations required for the phasing at the 3.5PN order
are done in harmonic coordinates.
Harmonic coordinates are preferred as computations
that lead to ready to use search templates
for compact binaries in quasi-circular orbits are usually
performed in harmonic gauge \cite{ref1NEW}.

This paper has the following plan.
In Sec.~\ref{phasing:PhasingSec},
we outline the procedure, detailed in Ref.~\cite{DGI}, 
to perform the phasing.
Section~\ref{phasing:analex3.5PNSec}       
provides explicit formulae 
required for the 3.5PN accurate phasing,
which also include 
3.5PN accurate equations
for the secular and periodic variations
of the orbital elements
involved in the phasing. 
The pictorial representation of the main results
and the related discussions are presented 
in Sec.~\ref{phasing:sec_visualization}.
Finally, in Sec.~\ref{phasing:sec_conclusions}, 
we give a brief summary and point out possible extensions. 
Appendices~\ref{phasing:Appendix:vmu} 
and \ref{phasing:Appendix:PNadiabaticSec} 
deal with computational details
and some supplementary results.


\section{Brief description of GW phasing for inspiralling eccentric binaries}

\label{phasing:PhasingSec}

In this section, 
we summarize the basic ideas of GW phasing, 
as detailed in Ref.~\cite{DGI}.
We also discuss 
what is required for this purpose
and how the method of variation of constants will be invoked.

\subsection{Basic structure of GW phasing}

Let us first describe the method,
developed in Ref.~\cite{DGI},
to implement the GW phasing 
for compact binaries in inspiralling eccentric orbits.
The theoretical templates
required by the GW interferometers 
consist of the two independent 
GW polarization states $h_{+}$ and $h_{\times}$.
The appropriate expressions for $h_{+}$ and $h_{\times}$,
expressed in terms of the binary's intrinsic dynamical
variables and location,
are given by
\begin{subequations}
\label{phasing_eq:1}
\begin{align}
\label{phasing_eq:1a}
h_{+} &
= \frac{1}{2} ( p_i p_j - q_i q_j ) h_{ij}^{\rm TT}
\,,
\\
\label{phasing_eq:1b}
h_{\times} & 
= \frac{1}{2} ( p_i q_j + p_j q_i ) h_{ij}^{\rm TT}
\,,
\end{align}
\end{subequations}
where $h_{ij}^{\rm TT}$ is the transverse-traceless (TT) part 
of the radiation field.
The two orthogonal unit vectors $\bm{p}$ and $\bm{q}$ span 
the plane of the sky, i.e.,
the plane transverse to the radial direction linking
the source to the observer.
   
The TT radiation field is given,
by the existing GW generation 
formalisms \cite{BDIWW,BIWW96,WW96,ref1NEW,ABIQ04}, 
as a PN expansion in $(v/c)$. 
In this paper, 
for simplicity, we will restrict
$h_{ij}^{\rm TT}$ to its leading ``quadrupolar'' order 
and denote it by $h_{ij}^{\rm TT}|_{\rm Q}$.
However, higher-PN corrections to $h_{ij}^{\rm TT}$ 
are available in the existing literature \cite{WW96,GI97}. 
The explicit expression for $h_{ij}^{\rm TT}|_{\rm Q}$,  
in terms of 
the relative separation vector $\bm{r}$ and
the relative velocity vector $\bm{v}$,
reads 
\begin{align}
\label{phasing_eq:2}
h_{km}^{\rm TT} \big|_{\rm Q} 
& = \frac{ 4 G \mu }{ c^4 R'} 
\mathcal{P}_{ijkm} (\bm{N})
\left( v_{i} v_{j} - \frac{ G M }{r} n_{i} n_{j} \right)
\,,
\end{align}
where $\mathcal{P}_{ijkm} (\bm{N})$
is the usual TT 
projection operator projecting normal to $\bm{N}$,
where $\bm{N} = \bm{R}' / R'$ is the line-of-sight unit vector 
from the binary to the observer, 
and $R' = |\bm{R}'|$ is the corresponding radial distance. 
The reduced mass of the binary $\mu$ is given
by $m_{1} m_{2} / M$,
where $M \equiv m_{1} + m_{2}$ 
is the total mass of the binary consisting 
of individual masses $m_{1}$ and $m_{2}$.
The components of the 
unit relative separation vector $\bm{n} = \bm{r} / r$, 
where $r = |\bm{r}|$,
and the 
velocity vector $\bm{v} = d \bm{r} / dt$
are denoted by $n_{i}$ and $v_{i}$, respectively.

In order to compute 
$h_{+} |_{\rm Q}$ and $h_{\times} |_{\rm Q}$,
the expressions for the GW polarization states 
when their amplitudes are restricted to the leading quadrupolar order, 
one needs to choose a convention for the direction and orientation 
of the orbital plane with respect to the plane of sky.
We follow the convention used in Refs.~\cite{DGI,WW96,BIWW96}
for choosing the orthonormal triad
$(\bm{p},\bm{q},\bm{N})$ 
--- $\bm{N}$ from the source to the observer
and $\bm{p}$ toward the correspondingly defined ascending node --- 
and use
$\bm{r} 
= r \cos\phi \bm{p} 
+ r \sin\phi ( \cos i \bm{q} + \sin i \bm{N} )$,
where
$i$ denotes the inclination angle of the orbital plane
with respect to the plane of the sky.
This leads to the following lowest-order contributions
to the two independent GW polarization states 
as functions of the relative separation $r$ 
and the true anomaly $\phi$, 
i.e., the polar angle of $\bm{r}$, 
and their time derivatives $\dot{r}$ and $\dot{\phi}$, 
which read \cite{DGI}
\begin{subequations}
\label{phasing_eq:3}
\begin{align}
\label{phasing_eq:3a}
h_{+} (r,\phi,\dot{r},\dot{\phi}) \big|_{\rm Q} &
= - \frac{G \mu}{c^4 R'}
\bigg\{
(1 + C^2) 
\bigg[ 
2 \dot{r} r \dot{\phi} \sin 2 \phi
\nonumber
\\
& \quad
+ \bigg( \frac{G M}{r} + r^2 \dot{\phi}^2 -\dot{r}^2 \bigg) \cos 2 \phi
\bigg]
\nonumber
\\
& \quad
+ S^2 \bigg( \frac{G M}{r} - r^2 \dot{\phi}^2 - \dot{r}^2 \bigg)
\bigg\}
\,,
\\
\label{phasing_eq:3b}
h_{\times} (r,\phi,\dot{r},\dot{\phi}) \big|_{\rm Q} &
= - \frac{2 G \mu C}{c^4 R'}
\bigg[
\bigg( \frac{G M}{r} + r^2 \dot{\phi}^2 - \dot{r}^2 \bigg) \sin 2 \phi
\nonumber
\\
& \quad
- 2 \dot{r} r \dot{\phi} \cos 2 \phi
\bigg]
\,,
\end{align}
\end{subequations}
where 
$C$ and $S$ are shorthand notations for
$\cos i$ and $\sin i$, respectively.
The orbital phase is denoted by 
$\phi$, 
$\dot{\phi} = d \phi / dt$, and 
$\dot{r} = dr / dt = \bm{n} \cdot \bm{v}$. 

In order to achieve the GW phasing
for compact binaries of arbitrary mass ratio
moving in inspiralling eccentric binaries,
we need to provide 
explicit expressions describing
the temporal evolution of 
$r(t)$, $\phi(t)$, $\dot{r}(t)$, and $\dot{\phi}(t)$,
the only dynamical variables appearing in Eqs.~\eqref{phasing_eq:3}.
Following Ref.~\cite{DGI}, 
we refer to, as \emph{phasing}, 
an explicit way to define the latter time dependences.
This is the crucial input
to derive ready to use waveforms $h_{+}(t)$ and $h_{\times}(t)$.
Adapting the formalism presented in Ref.~\cite{DGI},
we will provide the above desired time evolution
to the 3.5PN order in an almost parametric manner.

However, 
there are a few points that we want to clarify
before we describe the above procedure.
First, 
as explained in Ref.~\cite{DGI},
the possibility of obtaining explicit expressions 
for the GW polarizations $h_{+} |_{\rm Q}$ and $h_{\times} |_{\rm Q}$
in terms of the relative dynamics,
specified by $r(t)$, $\phi(t)$, $\dot{r}(t)$, and $\dot{\phi}(t)$,
relies on the possibility of going to
a suitable defined center-of-mass (COM) frame. 
Though such a frame exists only to the 3PN order \cite{BI02},
we will not worry about the associated ``recoil''
of the center of mass at the 3.5PN order.
This is because,
as demonstrated in Ref.~\cite{DGI},
the influence of the recoil on waveforms
appears at the 4PN order and
we may safely neglect it
for the present computation.

A second point,
again detailed in Ref.~\cite{DGI}, 
is that the GW polarizations 
$h_{+} |_{\rm Q}$ and $h_{\times} |_{\rm Q}$
\emph{only} depend on the temporal evolution of 
$r(t)$, $\phi(t)$, $\dot{r}(t)$, and $\dot{\phi}(t)$,
as we are dealing with nonspinning compact objects. 
In the presence of spin interactions, 
the orbital plane is no longer fixed in space and 
one needs to introduce 
further variables \cite{DT92,KG05_so_para,KG06_k_obs,DS88}.
However, expressions for 
$h_{+} |_{\rm Q}$ and $h_{\times} |_{\rm Q}$,
valid for \emph{spinning} compact binaries
are recently obtained in Ref.~\cite{KG05_so_para} 
and the phasing for such binaries 
will be reported elsewhere.

Finally, we want to discuss 
the choice of the underlying coordinate system.
Though the explicit functional forms of 
$h_{+} (r,\phi,\dot{r},\dot{\phi}) |_{\rm Q}$ and 
$h_{\times} (r,\phi,\dot{r},\dot{\phi}) |_{\rm Q}$,
as well as the phasing relations 
$r(t)$, $\phi(t)$, $\dot{r}(t)$, and $\dot{\phi}(t)$
depend on the used coordinate system, 
the final results 
$h_{+}(t)$ and $h_{\times}(t)$ do not.
Note that $h_{ij}^{\rm TT}$ and therefore 
$h_{+}(t)$ and $h_{\times}(t)$
are coordinate independent asymptotic quantities.
In this paper, we consistently work in 
harmonic coordinates because 
(i) they will allow us to write down 
explicit analytical expressions 
for the orbital phasing 
$r(t)$, $\phi(t)$, $\dot{r}(t)$, and $\dot{\phi}(t)$,
based on Ref.~\cite{MGS}, and
(ii) the harmonic coordinate systems are used 
in the standard GW generation formalisms
to derive the amplitude expressions,
giving higher-PN corrections to $h_{ij}^{\rm TT}$.

In the next subsection, 
we will explain
a version of the general Lagrange
method of variation of arbitrary constants, 
which was employed to compute, 
within general relativity, the orbital 
evolution of the Hulse-Taylor binary pulsar \cite{TD83,TD_F85},
and applied for the 2.5PN accurate phasing in Ref.~\cite{DGI}.


\subsection{Improved method of variation of constants
and its implementation at the 3.5PN order}

\label{phasing:MethodSec}

The method,
presented in Ref.~\cite{DGI},
begins by splitting the relative acceleration $\bm{\mathcal{A}}$ 
of the compact binary into two parts,
an integrable leading part $\bm{\mathcal{A}}_0$ 
and a perturbative part $\bm{\mathcal{A}}'$, 
as $\bm{\mathcal{A}} = \bm{\mathcal{A}}_0 + \bm{\mathcal{A}}'$.
The method first constructs the solution to
the ``unperturbed'' system, 
defined by
\begin{subequations}
\label{phasing_eq:unpert}
\begin{align}
\label{phasing_eq:4a}
\dot{\bm{r}} & = \bm{v}
\,,
\\
\label{phasing_eq:4b}
\dot{\bm{v}} & = \bm{\mathcal{A}}_0 (\bm{r},\bm{v}).
\end{align}
\end{subequations}
The solution to the exact system
\begin{subequations}
\label{phasing_eq:full}
\begin{align}
\label{phasing_eq:5a}
\dot{\bm{r}} & = \bm{v}
\,,
\\
\label{phasing_eq:5b}
\dot{\bm{v}} & = \bm{\mathcal{A}} (\bm{r},\bm{v})
\,,
\end{align}
\end{subequations}
is then obtained by \emph{varying the constants}
in the generic solution of the unperturbed system,
given by Eqs.~\eqref{phasing_eq:unpert}.

In this paper, 
we work to the 3.5PN order and therefore 
$\bm{\mathcal{A}}_0$ will be the acceleration at the 3PN order and
$\bm{\mathcal{A}}'$ contains the reactive 2.5PN and 3.5PN 
contributions.
The method assumes --- as is true for 
$\bm{\mathcal{A}}_{\rm 3PN}^{\rm conservative}$ ---
that the unperturbed system admits sufficiently many integrals of motion
to be integrable.
At the 3PN order with
$\bm{\mathcal{A}}_0 = \bm{\mathcal{A}}_{\rm 3PN}$,
we have four first integrals:
the 3PN accurate energy and
the 3PN accurate angular-momentum vector of the binary.
We denote these quantities, written in the 3PN accurate 
COM frame 
by $c_1$ and $c_2^i$:
\begin{subequations}
\label{phasing_eq:6}
\begin{align}
\label{phasing_eq:6a}
c_1 & 
= {E} (\bm{r}, \bm{v})
\big|_{\rm 3PN~COM}
\,,
\\
\label{phasing_eq:6b}
c_2^i & 
= {L}_i (\bm{r}, \bm{v})
\big|_{\rm 3PN~COM}
\,.
\end{align}
\end{subequations}

At the 3PN order,
the functional form of the solution
to the unperturbed equations of motion
in the COM frame,
following Refs.~\cite{TD83,MGS},
may be expressed as 
\begin{subequations}
\label{phasing_eq:7}
\begin{align}
\label{phasing_eq:7a}
r & = S(l; c_{1}, c_{2})
\,,
\\
\label{phasing_eq:7b}
\dot{r} & = n \frac{ \partial S }{ \partial l } (l; c_{1}, c_{2})
\,,
\\
\label{phasing_eq:7c}
\phi & = \lambda + W(l; c_{1}, c_{2})
\,,
\\
\label{phasing_eq:7d}
\dot{\phi} & = ( 1 + k ) n + n \frac{\partial W}{\partial l} (l; c_{1}, c_{2})
\,,
\end{align}
\end{subequations}
where $\lambda$ 
and $l$ are two basic angles, 
which are $2 \pi$ periodic,
and $c_2 = |c_2^i|$.
The functions $S(l)$ and $W(l)$ and hence their derivatives  
$\frac{\partial S}{\partial l} (l)$ and 
$\frac{\partial W}{\partial l} (l)$
are also periodic in $l$.
In the above equations,
$n$ denotes the unperturbed ``mean motion'',
given by $n = 2 \pi / P$,
$P$ being the radial (periastron to periastron) period,
while $k = \Delta \Phi / (2\pi)$,
where $\Delta \Phi$ represents the advance of periastron
in the time interval $P$.
The explicit 3PN accurate expressions for $P$ and $k$
in terms of $c_1$ and $c_2$ 
are obtainable from Refs.~\cite{MGS,DJS00B}.
The angles $l$ and $\lambda$ satisfy,
still for the unperturbed system,
$\dot{l} = n$ and $\dot{\lambda} = (1 + k) n$,
which integrate to
\begin{subequations}
\label{phasing_eq:8}
\begin{align}
\label{phasing_eq:8a}
l & = n (t - t_0) + c_l
\,,
\\
\label{phasing_eq:8b}
\lambda & = (1 + k) n (t - t_0) + c_\lambda
\,,
\end{align}
\end{subequations}
where $t_0$ is some initial instant and
the constants $c_l$ and $c_\lambda$ are the corresponding
values for $l$ and $\lambda$.
Note that 
the unperturbed solution depends
on four integration constants:
$c_1$, $c_2$, $c_l$, and $c_\lambda$.

Before
we explain
the prescription
to obtain the phasing at the 3.5PN order,
let us first present
our description of the 3PN accurate conservative dynamics.
The 3PN accurate parametric solution 
to the dynamics of compact binaries 
of arbitrary mass ratio
moving in eccentric orbits,
derived in Ref.~\cite{MGS},
allows us to describe the orbital motion
in the 3PN accurate COM frame.
The vectorial structure of $c_2^i$ [see Eq.~\eqref{phasing_eq:6b}],
indicates that
the unperturbed 3PN accurate motion takes place in a plane,
the so-called orbital plane.
The problem is restricted to a plane 
even in the presence of radiation reaction.
Therefore, we introduce polar coordinates 
in the orbital plane,  
$r$ and $\phi$, 
such that the relative separation vector $\bm{r}$
takes the following form
$\bm{r} = r \cos \phi \bm{i} + r \sin \phi \bm{j}$,
where we may choose 
$\bm{i} = \bm{p}$ and
$\bm{j} = \cos i \bm{q} + \sin i \bm{N}$ 
as the definition 
for the basic vectors. 
The dynamical variables
$r$ and $\phi$,
available in Ref.~\cite{MGS},
are parametrically given by 
\begin{subequations}
\label{phasing_eq:FinalParam3PNharmonic}
\begin{align}
r & = a_r \left( 1 - e_r \cos u \right)
\,,
\\
\phi - \phi_{0} & 
= (1 + k ) v 
+ \bigg( \frac{f_{4\phi}}{c^4} + \frac{f_{6\phi}}{c^6} \bigg) \sin 2v
\nonumber
\\
& \quad
+ \bigg( \frac{g_{4\phi}}{c^4} + \frac{g_{6\phi}}{c^6} \bigg) \sin 3v
+ \frac{ i_{6\phi} }{c^6} \sin 4v
\nonumber
\\
& \quad
+ \frac{ h_{6\phi} }{c^6} \sin 5v 
\,,
\\
\label{phasing_eq:9c}
\text{where} \quad
v & = 2 \arctan 
\left[ 
\left( \frac{ 1 + e_{\phi} }{ 1 - e_{\phi} } \right)^{1/2} 
\tan \frac{u}{2} 
\right]
\,.
\end{align}
\end{subequations}
In the above equations, 
$a_r$, $e_r$, and $e_{\phi}$ are
some 3PN accurate semimajor axis,
radial eccentricity, and 
angular eccentricity, respectively, while 
$f_{4\phi}$, $f_{6\phi}$, 
$g_{4\phi}$, $g_{6\phi}$, 
$i_{6\phi}$, and 
$h_{6\phi}$
are some PN accurate orbital functions.
The following 3PN accurate Kepler equation
connects the eccentric anomaly $u$ 
(and hence the true anomaly $v$)
to the coordinate time $t$
and reads 
\begin{align}
\label{phasing_eq:10}
l \equiv n \left( t - t_0 \right) 
& = u - e_t \sin u 
+ \bigg( \frac{g_{4t}}{c^4} + \frac{g_{6t}}{c^6} \bigg) (v - u)
\nonumber
\\
& \quad
+ \bigg( \frac{f_{4t}}{c^4} + \frac{f_{6t}}{c^6} \bigg) \sin v
+ \frac{i_{6t}}{c^6} \sin 2v
\nonumber
\\
& \quad
+ \frac{h_{6t}}{c^6} \sin 3v 
\,.
\end{align}
Here $e_t$ denotes some time eccentricity and
$g_{4t}$, $g_{6t}$, 
$f_{4t}$, $f_{6t}$, 
$i_{6t}$, and 
$h_{6t}$
are some PN accurate orbital functions.
All these PN accurate orbital elements and functions,
expressible in terms of $c_1$, $c_2$, and $\eta \equiv \mu / M$,
are available in Ref.~\cite{MGS}.

With the help of the above
3PN accurate parametric solution,
we write down explicit expressions for 
the functions $S(l)$ and $W(l)$:
\begin{subequations}
\label{phasing_eq:11}
\begin{align}
\label{phasing_eq:11a}
S(l;c_{1},c_{2}) & = a_r ( 1 - e_r \cos u )
\,,
\\
\label{phasing_eq:11b}
W(l;c_{1},c_{2}) & 
= ( 1 + k ) ( v - l )
+ \bigg( \frac{ f_{4\phi} }{c^4} + \frac{ f_{6\phi} }{c^6} \bigg) \sin 2 v
\nonumber
\\
& \quad
+ \bigg( \frac{ g_{4\phi} }{c^4} + \frac{ g_{6\phi} }{c^6} \bigg) \sin 3 v
+ \frac{ i_{6\phi} }{c^6} \sin 4 v
\nonumber
\\
& \quad
+ \frac{ h_{6\phi} }{c^6} \sin 5 v
\,.
\end{align}
\end{subequations}
We emphasize that for the phasing
the anomalies $v$ and $u$ 
in Eqs.~\eqref{phasing_eq:11} 
have to be expressed 
as functions of $l$, $c_1$, and $c_2$.
This can be achieved 
in the following way,
written symbolically as
$v = \mathcal{V} (l;c_1,c_2) 
= V[ \mathcal{U} (l;c_1,c_2)]$
and $u = \mathcal{U} (l;c_1,c_2)$.
First, recall that the function $v \equiv V(u)$ 
is defined by Eq.~\eqref{phasing_eq:9c}, and
second, the function $u = \mathcal{U} (l)$ is defined
by inverting the Kepler equation $l = l(u)$, 
given by Eq.~\eqref{phasing_eq:10}.
Finally, the function $v = \mathcal{V} (l)$
is obtained by inserting $u = \mathcal{U} (l)$
in $v = V(u)$, i.e.,
$v = \mathcal{V} (l) \equiv V[\mathcal{U} (l)]$.

In our computations, 
we use the following exact relation for $v - u$, 
which is also periodic in $u$:
\begin{align}
\label{phasing_eq:12}
v - u & = 2 \tan^{-1}
\left(
\frac{ \beta_{\phi} \sin u }{ 1 - \beta_{\phi} \cos u }
\right)
\,,
\end{align}
where $\beta_{\phi} = ( 1 - \sqrt{ 1 - e_\phi^2 } ) / e_\phi$
[for the derivation of Eq.~\eqref{phasing_eq:12},
see Appendix~\ref{phasing:Appendix:vmu}]. 

In line with the
method of variation of constants,
we write down the following general solution to the 
reactive 3.5PN accurate dynamics, Eqs.~\eqref{phasing_eq:full}, as
\begin{subequations}
\label{phasing_eq:13}
\begin{align}
\label{phasing_eq:13a}
r & = S(l;c_1,c_2)
\,,
\\
\label{phasing_eq:13b}
\dot{r} & = n \frac{\partial S}{\partial l} (l;c_1,c_2)
\,,
\\
\label{phasing_eq:13c}
\phi & = \lambda + W(l;c_1,c_2)
\,,
\\
\label{phasing_eq:13d}
\dot{\phi} & = ( 1 + k ) n + n \frac{\partial W}{\partial l} (l;c_1,c_2)
\,.
\end{align}
\end{subequations}
However,
the constants $c_1$ and $c_2$, appearing in Eqs.~\eqref{phasing_eq:7},
are now functions of time in Eqs.~\eqref{phasing_eq:13}: 
$c_1 = c_1 (t)$ and $c_2 = c_2 (t)$.
The temporal variation of the basic angles $l$ and $\lambda$,
entering Eqs.~\eqref{phasing_eq:13},
is now given by
\begin{subequations}
\label{phasing_eq:14}
\begin{align}
\label{phasing_eq:14a}
l & \equiv
\int_{t_0}^t n dt + c_l (t)
\,,
\\
\label{phasing_eq:14b}
\lambda & \equiv
\int_{t_0}^t ( 1 + k) n dt + c_\lambda (t)
\,,
\end{align}
\end{subequations}
involving two new temporally evolving quantities 
$c_l(t)$ and $c_\lambda (t)$.
In the method of variation of constants,
we search
for solutions of the exact system,
Eqs.~\eqref{phasing_eq:full}, in the form given by Eqs.~\eqref{phasing_eq:13}
and \eqref{phasing_eq:14} with four ``varying constants'':
$c_{1}(t)$, $c_{2}(t)$, $c_{l}(t)$, and $c_{\lambda}(t)$.
These four new variables 
replace the original four dynamical variables
$r(t)$, $\dot{r}(t)$, $\phi(t)$, and $\dot{\phi}(t)$,
and satisfy,
like the original phase-space variables,
first-order evolution equations \cite{TD83,TD_F85}.
These first-order evolution equations for
$c_{1}(t)$, $c_{2}(t)$, $c_{l}(t)$, and $c_{\lambda}(t)$
are available in Refs.~\cite{DGI,TD_F85}
and read in our notation
\begin{subequations}
\label{phasing_eq:15}
\begin{align}
\label{phasing_eq:15a}
\frac{d c_1}{dt} & =
\frac{ \partial c_1 (\bm{r},\bm{v}) }{ \partial v^i } {\mathcal{A}'}^{i}
\,,
\\
\label{phasing_eq:15b}
\frac{d c_2}{dt} & =
\frac{ \partial c_2 (\bm{r},\bm{v}) }{ \partial v^j } {\mathcal{A}'}^{j}
\,,
\\
\label{phasing_eq:15c}
\frac{d c_l}{dt} & =
-\left(
\frac{\partial S}{\partial l}
\right)^{-1} 
\left( 
\frac{\partial S}{\partial c_1} \frac{d c_1}{dt}
+\frac{\partial S}{\partial c_2} \frac{d c_2}{dt} 
\right)
\,,
\\
\label{phasing_eq:15d}
\frac{d c_\lambda}{dt} & =
-\frac{\partial W}{\partial l} \frac{d c_l}{dt} 
-\frac{\partial W}{\partial c_1} \frac{d c_1}{dt} 
-\frac{\partial W}{\partial c_2} \frac{d c_2}{dt}
\,.
\end{align}
\end{subequations}
In addition,
an alternative expression for $d c_l / dt$ 
is presented in Refs.~\cite{DGI,TD_F85}, 
which we used for some supplementary checks.

We observe that the definition of the sole angle $l$,
appearing in Eqs.~\eqref{phasing_eq:15},
given by
$l = \int_{t_0}^t n[c_{a} (t)] dt + c_{l} (t)$,
where $c_a$, $a = 1$, $2$ stands for $c_1$ and $c_2$,
is equivalent to the differential form
$dl/dt = n(c_{a}) + d c_l /dt = n(c_{a}) + F_l (l;c_{a})$.
This allowed Ref.~\cite{DGI}
to write the above set of equations,
namely, Eqs.~\eqref{phasing_eq:15} 
for $c_\alpha$ as functions of $t$,
as a set of differential equations 
for $c_\alpha$ as functions of $l$, 
which symbolically reads
\begin{align}
\label{phasing_eq:16}
\frac{ d c_\alpha }{dl} & =
\frac{ F_\alpha (l;c_a) }{ n(c_a) + F_l (l;c_a) }
\,,
\end{align}
where $\alpha = 1$, $2$, $l$, $\lambda$ and $a = 1$, $2$. 
Neglecting terms quadratic in $F_\alpha$, i.e.,
quadratic in the perturbation $\bm{\mathcal{A}}'$, 
terms of $\mathcal{O} (c^{-10})$ (and higher-PN orders), 
we can simplify the system above to 
\begin{align}
\label{phasing_eq:17}
\frac{ d c_\alpha}{dl} &
\simeq \frac{ F_\alpha (l;c_{a}) }{ n(c_{a}) } \equiv G_\alpha (l; c_a)
\,.
\end{align}
It is important to note at this stage
that in the above prescription,
which neglects $\mathcal{O} (c^{-10})$ terms,
the right-hand side of Eq.~\eqref{phasing_eq:17},
first, is \emph{just} a function of 
$c_1$, $c_2$, and the sole angle $l$ (and not of $\lambda$),
and second, it is a \emph{periodic} function of $l$.
This periodicity, 
together with the slow evolution of the $c_\alpha$'s,
clearly indicates
that the evolution of $c_\alpha (l)$ contains 
not only a slow secular drift, 
but also fast periodic oscillations. 
The secular drift occurs in the radiation-reaction time scale,
while the time scale for the periodic oscillations
is that of the orbital motion. 

Following Ref.~\cite{DGI},
we model the combination of 
the slow drift and the fast oscillations in $c_\alpha (l)$,
by a two-scale decomposition of $c_\alpha (l)$,
which reads
\begin{align}
\label{phasing_eq:18}
c_\alpha (l) & =
\bar{c}_\alpha (l) + \tilde{c}_\alpha (l)
\,.
\end{align}
In the above equation,
$\bar{c}_\alpha (l)$ denotes the slow drift,
which accumulates over the radiation-reaction time scale
to induce large changes in $c_\alpha (l)$, while
the fast oscillations in $c_\alpha (l)$
are denoted by $\tilde{c}_\alpha (l)$, 
which will be always smaller than $\bar{c}_\alpha (l)$.

In this paper,
we are interested in the 3.5PN accurate evolution
of $\bar{c}_\alpha (l)$ and $\tilde{c}_\alpha (l)$.
The corresponding evolution equations for 
$\bar{c}_\alpha (l)$ and $\tilde{c}_\alpha (l)$ 
follow from the combination of
Eqs.~(41), (43), and (44) of Ref.~\cite{DGI}
and read
\begin{subequations}
\label{phasing_eq:19}
\begin{align}
\label{phasing_eq:19a}
\frac{ d \bar{c}_\alpha }{dl} &
= \bar{G}_\alpha ( \bar{c}_a ) 
= \frac{1}{2 \pi} \int_0^{2 \pi} G(l; \bar{c}_a ) \, dl
\,,
\\
\label{phasing_eq:19b}
\frac{ d \tilde{c}_\alpha }{dl} &
= \tilde{G}_\alpha (l; \bar{c}_a ) 
= G_\alpha (l; \bar{c}_a ) - \bar{G}_\alpha ( \bar{c}_a )
\,.
\end{align}
\end{subequations}
Finally, the arguments
that gave the 2.5PN accurate unique zero-average expressions 
for $\tilde{c}_\alpha (l)$ in Ref.~\cite{DGI},
can be extended to obtain the 3.5PN accurate expressions 
for $\tilde{c}_\alpha (l)$.
This leads to the solution of Eq.~\eqref{phasing_eq:19b}, 
considered for fixed values of $\bar{c}_a$,
\begin{align}
\label{phasing_eq:20}
\tilde{c}_\alpha(l) & = 
\left[ 
\int \tilde{G}_\alpha (l;\bar{c}_a) \, dl
\right]_{\bar{c}_a = \bar{c}_a (l)}
= \int \tilde{F}_\alpha (l;\bar{c}_a) \frac{dl}{n}
\,.
\end{align}
We recall from Ref.~\cite{DGI}
that the indefinite integral in Eq.~\eqref{phasing_eq:20}
is defined as the unique zero-average 
periodic primitive of the zero-average (periodic)
function $\tilde{G}_\alpha (l)$.
During that integration, the arguments $\bar{c}_a$
are kept fixed, and, after the integration, they are replaced 
by the slowly drifting solution of Eq.~\eqref{phasing_eq:19a}.

Furthermore, 
a consistency check of the two-scale method
was also provided in Ref.~\cite{DGI} 
by exploring the effects 
of the above neglected second-order terms
in Eqs.~\eqref{phasing_eq:17}--\eqref{phasing_eq:20}.
It was shown there
that the separation between the two scales
remains valid on very long time scales and 
that second-order (and higher-order) effects cause only
fractionally small
separate corrections to the evolution of
$\bar{c}_\alpha (l)$ and $\tilde{c}_\alpha (l)$.

In the next section,
we will implement the arguments developed above
and obtain explicit expressions
to perform the GW phasing at the 3.5PN order.


\section{Analytical expressions for the 3.5PN accurate phasing}

\label{phasing:analex3.5PNSec}

In this section,
we apply the above described 
improved method of variation of arbitrary constants, 
which gave us the evolution equations for
$\bar{c}_\alpha$ and $\tilde{c}_\alpha$, 
to achieve the GW phasing in the following way.
First, 
we compute the 3PN accurate parametric expressions for
the dynamical variables $r$, $\dot{r}$, $\phi$, and $\dot{\phi}$
entering the expressions for 
$h_{+} |_{\rm Q}$ and $h_{\times} |_{\rm Q}$,
given by Eqs.~\eqref{phasing_eq:3}.
Then we solve the evolution equations for
$c_{1}$, $c_{2}$, $c_{l}$, and $c_{\lambda}$, 
given by Eqs.~\eqref{phasing_eq:17},
on the 3PN accurate orbital dynamics, 
given in Eqs.~\eqref{phasing_eq:7}.
This leads to an evolution system,
given by Eqs.~\eqref{phasing_eq:19},
where the right-hand side contains 
dominant $\mathcal{O} (c^{-5})$ terms and their first corrections,
i.e., $\mathcal{O} (c^{-7})$ terms.
Later,
we will impose these variations
on the 3PN accurate expressions 
for the dynamical variables 
$r$, $\dot{r}$, $\phi$, and $\dot{\phi}$,
appearing in $h_{+} |_{\rm Q}$ and $h_{\times} |_{\rm Q}$, 
given by Eqs.~\eqref{phasing_eq:3}.
This allows us to obtain GW polarizations,
which are Newtonian accurate in their amplitudes and 
3.5PN accurate in the orbital dynamics.
Following Ref.~\cite{DGI}, 
the above procedure is called
3.5PN accurate phasing of gravitational waves.
In our computations,
the quasi-periodic oscillations in $c_\alpha$,
governed by $\tilde{G}_\alpha$,
are restricted to the 1PN reactive order.
Therefore,
we will not explore higher-PN corrections 
to the above $\tilde{G}_\alpha$.
However, 
in Appendix~\ref{phasing:Appendix:PNadiabaticSec},
we will present the consequences
of considering higher-PN corrections to $\bar{G}_\alpha$ by computing
$\mathcal{O} (c^{-9})$ contributions to relevant
$d \bar{c}_\alpha /dt$.
This is desirable as $\bar{G}_\alpha$ directly contributes to the
highly important adiabatic evolution of $h_+$ and $h_\times$.

In this paper, 
we follow Ref.~\cite{DGI}    
and employ as $c_1$ the mean motion $n$, 
and as $c_2$ the time eccentricity $e_t$,
instead of the energy $E$ and the angular momentum $L$, respectively,
to describe the PN accurate reactive dynamics.
This can be done by employing
3PN accurate expressions for $n$ and $e_t$
in terms of $E$ and $L$,
derived in Ref.~\cite{MGS}.
This implies that 
first, we have to express 
the 3PN accurate orbital dynamics
in terms of $l$, $n$, and $e_t$.
Second, using $n$ and $e_t$, 
instead of $E$ and $L$ 
as $c_1$ and $c_2$, respectively,
we need to derive the evolution equations for 
$d n / dt$, 
$d e_t / dt$, 
$d c_l / dt$, and 
$d c_\lambda / dt$
in terms of $l$, $n$, and $e_t$. 
This will follow straightforwardly from Eqs.~\eqref{phasing_eq:15}.
Using these expressions, the evolution equations,
namely, Eqs.~\eqref{phasing_eq:19}, for
$\bar{n}$, $\bar{e}_t$, $\bar{c}_l$, $\bar{c}_\lambda$, 
$\tilde{n}$, $\tilde{e}_t$, $\tilde{c}_l$, and $\tilde{c}_\lambda$ 
will be obtained in terms of $l$, $\bar{n}$, and $\bar{e}_t$.

Let us now obtain the explicit 3PN accurate
expressions for the orbital dynamics.


\subsection{3PN accurate conservative dynamics}

As mentioned earlier, we restrict in this paper
the conservative dynamics to the 3PN order.
Below, we present the 3PN accurate orbital dynamics,
namely, $r$, $\dot{r}$, $\phi$, and $\dot{\phi}$,
as given by Eqs.~\eqref{phasing_eq:13}, 
explicitly in terms of $u$, $n$, and $e_t$.
This straightforward computation employs explicit expressions 
for the PN orbital elements 
$a_r$, $e_r$, $e_\phi$, $k$,
$f_{4\phi}$, $f_{6\phi}$, 
$g_{4\phi}$, $g_{6\phi}$, 
$i_{6\phi}$, $h_{6\phi}$,
$g_{4t}$, $g_{6t}$, 
$f_{4t}$, $f_{6t}$, 
$i_{6t}$, and $h_{6t}$	
of the generalized quasi-Keplerian representation,
available in Ref.~\cite{MGS}.  
We note that the above PN orbital elements,
given in terms of $E$ and $L$ in Ref.~\cite{MGS},
can easily be expressed in terms of $n$ and $e_t$
with the help of the following 3PN accurate relations 
for $- 2 E$ and $- 2 E L^2$:
\begin{widetext}
\begin{subequations}
\label{phasing_eq:21}
\begin{align}
- 2 E & = (G M n)^{2/3} 
\bigg \{
1
+ \frac{ \xi ^{2/3} }{12} ( 15 - \eta )
+ \frac{ \xi ^{4/3} }{24} 
\bigg [
15 - 15 \eta - \eta^2 + \frac{ 24 }{ \sqrt{ 1 - e_t^2 } } ( 5 - 2 \eta )
\bigg ]
+ \frac{ \xi^2 }{5184} 
\bigg [
- 4995 - 6075 \eta - 450 \eta^2 
\nonumber
\\
& \quad
- 35 \eta^3 
+ \frac{ 864 }{ \sqrt{ 1 - e_t^2 } } ( 15 + 23 \eta - 20 \eta^2 )
+ \frac{ 18 }{ (1 - e_t^2)^{3/2} } 
\left( 
11520 - 15968 \eta + 123 \pi^2 \eta + 2016 \eta^2 
\right)
\bigg ]
\bigg \}
\,, 
\\
- 2 E L^2 & = ( 1 - e_t^2 ) 
\biggl (
1
+ \frac{ \xi^{2/3} }{ 4 ( 1 - e_t^2 ) } 
\left[ 9 + \eta - ( 17 - 7 \eta ) e_t^2 \right] 
+ \frac{ \xi^{4/3} }{24 ( 1 - e_t^2 )^2 }
\bigg[ 
189 - 45 \eta + \eta^2 
- 2 ( 111 + 7 \eta + 15 \eta^2 ) e_t^2
\nonumber
\\
& \quad
+ ( 225 - 277 \eta + 29 \eta^2 ) e_t^4
- ( 360 - 144 \eta ) e_t^2 \sqrt{1 - e_t^2} 
\bigg] 
+ \frac{ \xi^2 }{ 6720 ( 1-e_t^2 )^3 }
\biggl \{
35 ( 5535 - 9061 \eta + 246 \pi^2 \eta 
\nonumber
\\
& \quad
+ 142 \eta^2 - \eta^3 ) 
+ ( 
299145 - 1197667 \eta + 25830 \pi^2 \eta + 173250 \eta^2 + 2345 \eta^3 
) e_t^2
+ 35 ( 3549 - 12783 \eta + 6154 \eta^2
\nonumber
\\
& \quad
- 131 \eta^3 ) e_t^4
- 35 ( 2271 - 7381 \eta + 2414 \eta^2 - 65 \eta ^3 ) e_t^6
+ 70 \left[ 
24 ( 45 - 13 \eta - 2 \eta^2 )
- ( 17880 - 20120 \eta 
\right.
\nonumber
\\
& \quad
\left.
+ 123 \pi^2 \eta + 2256 \eta^2 ) e_t^2
+ 96 ( 55 - 40 \eta + 3 \eta^2 ) e_t^4
\right] \sqrt{1-e_t^2} 
\biggr \}
\biggr ) 
\,, 
\end{align}
\end{subequations}
where $\xi \equiv G M n / c^3$ and 
$\eta \equiv \mu / M = m_{1} m_{2} / ( m_{1} + m_{2} )^2$.
These two relations 
follow from inverting the
3PN accurate relations for 
the orbital period $P = 2 \pi / n$ 
and the squared time eccentricity $e_t^2$ 
in terms of $E$ and $L$ 
presented in Eqs.~(25c) and (25d) in Ref.~\cite{MGS}.

In addition, 
to compute expressions for $\dot{r}$ and $\dot{\phi}$,
we use the following relations:
\begin{subequations}
\label{phasing_eq:22}
\begin{align}
\label{phasing_eq:22a}
\frac{ \partial S }{\partial l} &
= a_r e_r \sin u \frac{ \partial u }{\partial l}
\,,
\\
\label{phasing_eq:22b}
\frac{ \partial W }{\partial l} &
= \left[
1 + k 
+ 2 \bigg( \frac{ f_{4\phi} }{c^4} + \frac{ f_{6\phi} }{c^6} \bigg) \cos 2 v
+ 3 \bigg( \frac{ g_{4\phi} }{c^4} + \frac{ g_{6\phi} }{c^6} \bigg) \cos 3 v
+ 4 \frac{ i_{6\phi} }{c^6} \cos 4 v
+ 5 \frac{ h_{6\phi} }{c^6} \cos 5 v
\right]
\frac{ \partial v }{\partial u}
\frac{ \partial u }{\partial l}
- ( 1 + k )
\,,
\\
\label{phasing_eq:22c}
\frac{ \partial u }{\partial l} &
= \bigg \{ 
1 - e_t \cos u 
- \frac{ g_{4t} }{c^4} - \frac{ g_{6t} }{c^6}
+ \bigg[
\frac{ g_{4t} }{c^4} + \frac{ g_{6t} }{c^6}
+ \bigg( \frac{ f_{4t} }{c^4} + \frac{ f_{6t} }{c^6} \bigg) \cos v
+ 2 \frac{ i_{6t} }{c^6} \cos 2 v
+ 3 \frac{ h_{6t} }{c^6} \cos 3 v
\bigg]
\frac{ \partial v }{\partial u} 
\bigg \}^{-1}
\,,
\\
\label{phasing_eq:22d}
\frac{ \partial v }{\partial u} &
= \frac{ ( 1 - e_\phi^2 )^{1/2} }{ 1 - e_\phi \cos u}
\,. 
\end{align}
\end{subequations}  
The radial motion, 
defined by $r(l,n,e_t)$ and $\dot{r}(l,n,e_t)$,
reads (both in the compact and in the 3PN expanded form)
\begin{subequations}
\label{phasing_eq:23}
\begin{align}
\label{phasing_eq:23a}
r & = S (l,n,e_t) = a_r (n,e_t) [ 1 - e_r (n,e_t) \cos u ]
= r_{\rm N} 
+ r_{\rm 1PN} 
+ r_{\rm 2PN} 
+ r_{\rm 3PN}
\,, 
\end{align}
where
\begin{align}
r_{\rm N} & = 
\left( \frac{G M}{n^2} \right)^{1/3} ( 1 - e_t \cos u )
\,,
\\
r_{\rm 1PN} & = 
r_{\rm N} \times
\frac{ \xi ^{2/3} }{ 6 ( 1 - e_t \cos u ) }
[ - 18 + 2 \eta - ( 6 - 7 \eta ) e_t \cos u ]
\,,
\\
r_{\rm 2PN} & =
r_{\rm N} \times
\frac{ \xi^{4/3} }{ 72 (1 - e_t^2 ) (1 - e_t \cos u) }
\Bigl \{
- 72 (4 - 7 \eta ) 
+ \left[ 
72 + 30 \eta + 8 \eta^2 - ( 72 - 231 \eta + 35 \eta ^2 ) e_t \cos u
\right] (1 - e_t^2 ) 
\nonumber
\\
& \quad
- 36 ( 5 - 2 \eta ) (2 + e_t \cos u ) \sqrt{1 - e_t^2} 
\Bigr \} 
\,,
\\
r_{\rm 3PN} & = 
r_{\rm N} \times
\frac{ \xi^2 }{ (1 - e_t^2 )^2 (1 - e_t \cos u) }
\biggl (
- \frac{70}{3} 
+ \frac{56221}{840} \eta - \frac{123}{64} \pi^2 \eta
- \frac{151}{36} \eta^2
+ \frac{2}{81} \eta^3 
+ \left(
- \frac{2}{3}
+ \frac{87}{16} \eta 
- \frac{437}{144} \eta^2
+ \frac{49}{1296} \eta^3 
\right) \nonumber
\\
& \quad
\times e_t \cos u
+ \left[
- \frac{52}{3} 
+ \frac{2099}{35} \eta - \frac{41}{64} \pi^2 \eta 
- \frac{341}{18} \eta^2
- \frac{4}{81} \eta^3
+ \left(
\frac{4}{3}
- \frac{87}{8} \eta 
+ \frac{437}{72} \eta^2
- \frac{49}{648} \eta^3 
\right) e_t \cos u
\right] e_t^2
\nonumber
\\
& \quad
+ \left[
\frac{2}{3} 
- \frac{1}{8} \eta
+ \frac{5}{36} \eta^2 
+ \frac{2}{81} \eta^3
+ \left( 
- \frac{2}{3}
+ \frac{87}{16} \eta 
- \frac{437}{144} \eta^2
+ \frac{49}{1296} \eta^3 
\right) e_t \cos u
\right] e_t^4
+ \biggl \{
- 30 
+ \frac{412}{9} \eta - \frac{41}{96} \pi^2 \eta
- \frac{10}{3} \eta^2
\nonumber
\\
& \quad
+ \left(
- \frac{45}{2} 
+ \frac{1247}{36} \eta - \frac{41}{192} \pi^2 \eta
- \frac{31}{6} \eta^2
\right) e_t \cos u 
+ \left[ 
- 10
+ \frac{29}{3} \eta
- \frac{11}{3} \eta^2
+ \left( 
\frac{5}{2} 
- \frac{83}{12} \eta 
+ \frac{5}{3} \eta^2 
\right) e_t \cos u
\right] e_t^2
\biggr \} 
\nonumber
\\
& \quad
\times
\sqrt{1 - e_t^2} 
\biggr )
\,, 
\end{align}
\end{subequations}
and
\begin{subequations}
\label{phasing_eq:24}
\begin{align}
\label{phasing_eq:24a}
\dot{r} & = n \frac{ \partial S }{ \partial l } (l,n, e_t)
= \dot{r}_{\rm N} 
+ \dot{r}_{\rm 1PN} 
+ \dot{r}_{\rm 2PN} 
+ \dot{r}_{\rm 3PN}
\,,
\end{align}
where
\begin{align}
\dot{r}_{\rm N} & = 
\frac{ (G M n)^{1/3} }{ (1 - e_t \cos u) } e_t \sin u
\,,
\\
\dot{r}_{\rm 1PN} & = 
\dot{r}_{\rm N} \times 
\frac{ \xi^{2/3} }{6} ( 6 - 7\eta )
\,,
\\
\dot{r}_{\rm 2PN} & = 
\dot{r}_{\rm N} \times 
\frac{ \xi^{4/3} }{ 72 ( 1 - e_t \cos u)^3 }
\biggl [ 
- 468 - 15 \eta + 35 \eta^2
+ ( 135 \eta - 9 \eta^2 ) e_t^2 
+ ( 324 + 342 \eta - 96 \eta^2 ) e_t \cos u 
+ ( 216 - 693 \eta 
\nonumber
\\
& \quad
+ 105 \eta^2 ) ( e_t \cos u )^2
- ( 72 - 231 \eta + 35 \eta^2 ) ( e_t \cos u )^3
+ \frac{ 36 }{ \sqrt{1 - e_t^2} } 
( 1 - e_t \cos u )^2 ( 4 - e_t \cos u ) ( 5 - 2 \eta )
\biggr ]
\,,
\\
\dot{r}_{\rm 3PN} & = 
\dot{r}_{\rm N} \times 
\frac{ \xi ^2 }{ ( 1 - e_t^2 )^{3/2} ( 1 - e_t \cos u)^5 }
\Biggl ( 
75
- \frac{2071}{18} \eta + \frac{41}{48} \pi^2 \eta
+ \frac{41}{3} \eta^2
+ \left( 5 + \frac{25}{6} \eta + \frac{1}{3} \eta^2  \right) e_t^2 
+ \left[
- \frac{645}{2}
+ \frac{17815}{36} \eta 
\right.
\nonumber
\\
& \quad
\left.
- \frac{697}{192} \pi^2 \eta
- \frac{359}{6} \eta^2
- \left( \frac{35}{2} + \frac{283}{12} \eta - \frac{1}{3} \eta^2 \right) e_t^2
\right] e_t \cos u
+ \left[
540
- \frac{7460}{9} \eta + \frac{287}{48} \pi^2 \eta
+ \frac{308}{3} \eta^2
+ \left( 20 + \frac{158}{3} \eta 
\right. \right.
\nonumber
\\
& \quad
\left. \left.
- \frac{14}{3} \eta^2 \right) e_t^2
\right] ( e_t \cos u )^2
+ \left[
- 435
+ \frac{12025}{18} \eta - \frac{451}{96} \pi^2 \eta
- \frac{257}{3} \eta^2
- \left( 5 + \frac{349}{6} \eta - \frac{26}{3} \eta^2 \right) e_t^2
\right] ( e_t \cos u )^3
\nonumber
\\
& \quad
+ \left[
165
- \frac{4565}{18} \eta + \frac{41}{24} \pi^2 \eta
+ \frac{103}{3} \eta^2
- \left( 5 - \frac{191}{6} \eta + \frac{19}{3} \eta^2 \right) e_t^2
\right] ( e_t \cos u )^4
+ \left[
- \frac{45}{2}
+ \frac{1247}{36} \eta - \frac{41}{192} \pi^2 \eta
- \frac{31}{6} \eta^2
\right.
\nonumber
\\
& \quad
\left.
+ \left( \frac{5}{2} - \frac{83}{12} \eta  + \frac{5}{3} \eta^2 \right) e_t^2
\right] ( e_t \cos u )^5
+ \biggl \{
- \frac{311}{6}
+ \frac{3599}{48} \eta 
- \frac{41}{64} \pi^2 \eta
- \frac{787}{144} \eta^2
- \frac{49}{1296} \eta^3
+ \left(
\frac{191}{6}
+ \frac{1189}{35} \eta 
\right.
\nonumber
\\
& \quad
\left.
- \frac{41}{64} \pi^2 \eta
- \frac{1591}{72} \eta^2
+ \frac{52}{81} \eta^3
\right) e_t^2
+ \left( 
- 47 \eta
+ \frac{165}{8} \eta^2 
- \frac{17}{12} \eta^3 
\right) e_t^4
+ \left( 
\frac{23}{16} \eta
- \frac{73}{16} \eta^2 
+ \frac{13}{16} \eta^3
\right) e_t^6
+ \left[
\frac{685}{6}
- \frac{60997}{280} \eta 
\right.
\nonumber
\\
& \quad
\left.
+ \frac{41}{16} \pi^2 \eta
+ \frac{170}{9} \eta^2
+ \frac{55}{648} \eta^3
+ \left(
- \frac{325}{6}
- \frac{10083}{560} \eta + \frac{41 }{32} \pi^2 \eta
+ \frac{3769}{144} \eta^2
+ \frac{673}{1296} \eta^3 
\right) e_t^2
+ \left( 
\frac{731}{16} \eta 
- \frac{169}{16} \eta^2 
\right. \right.
\nonumber
\\
& \quad
\left. \left.
- \frac{29}{48} \eta^3  
\right) e_t^4
\right] e_t \cos u
+ \left[
- \frac{425}{6} 
+ \frac{10531}{70} \eta - \frac{205}{64} \pi^2 \eta 
+ \frac{85}{18} \eta^2
- \frac{379}{324} \eta^3
+ \left(
\frac{65}{6}
+ \frac{5981}{140} \eta - \frac{41}{64} \pi^2 \eta 
- \frac{3085}{72} \eta^2
\right. \right.
\nonumber
\\
& \quad
\left. \left.
+ \frac{257}{162} \eta^3
\right) e_t^2
+ \left( 
- 3 \eta 
+ \frac{29}{8} \eta^2 
- \frac{5}{12} \eta^3 
\right) e_t^4
\right] ( e_t \cos u )^2 
+ \left[
\frac{35}{6}
+ \frac{10699}{840} \eta + \frac{41}{32} \pi^2 \eta
- \frac{463}{18} \eta^2
+ \frac{299}{648} \eta^3 
\right.
\nonumber
\\
& \quad
\left.
+ \left( 
\frac{85}{6}
- \frac{609}{8} \eta 
+ \frac{335}{9} \eta^2
- \frac{299}{648} \eta^3 
\right) e_t^2 
\right] ( e_t \cos u )^3 
+ \left[
\frac{10}{3} 
- \frac{435}{16} \eta
+ \frac{2185}{144} \eta^2
- \frac{245}{1296} \eta^3
+ \left(
- \frac{10}{3}
+ \frac{435}{16} \eta 
\right. \right.
\nonumber
\\
& \quad
\left. \left.
- \frac{2185}{144} \eta^2
+ \frac{245}{1296} \eta^3 
\right) e_t^2
\right] ( e_t \cos u )^4 
+ \left[ 
- \frac{2}{3}
+ \frac{87}{16} \eta
- \frac{437}{144} \eta^2
+ \frac{49}{1296} \eta^3
+ \left(
\frac{2}{3}
- \frac{87}{16} \eta 
+ \frac{437}{144} \eta^2
- \frac{49}{1296} \eta^3 
\right) e_t^2
\right] 
\nonumber
\\
& \quad
\times ( e_t \cos u )^5
\biggr \}
\sqrt{1 - e_t^2}
\Biggr )
\,. 
\end{align}
\end{subequations}

The angular motion, 
described in terms of $\phi$ and $\dot{\phi}$,
is given by
\begin{subequations}
\label{phasing_eq:25}
\begin{align}
\label{phasing_eq:25a}
\phi(\lambda,l) & = \lambda + W(l)
\,,
\\
\label{phasing_eq:25b}
\lambda & = ( 1 + k ) l
\,,
\\
\label{phasing_eq:25c}
W(l) & 
= W_{\rm N} 
+ W_{\rm 1PN} 
+ W_{\rm 2PN} 
+ W_{\rm 3PN}
\,,
\end{align}
where
\begin{align}
k & 
= \frac{ 3 \xi ^{2/3} }{ 1 - e_t^2 }
+ \frac{ \xi ^{4/3} }{ 4 ( 1 - e_t^2 )^2 } 
\left[ 78  - 28 \eta + ( 51 - 26 \eta ) e_t^2 \right]
+ \frac{ \xi^2 }{ 128 ( 1 - e_t^2 )^3 }
\Big \{ 
18240 - 25376 \eta + 492 \pi^2 \eta + 896 \eta ^2 
\nonumber
\\
& \quad
+ ( 28128 - 27840 \eta + 123 \pi^2 \eta + 5120 \eta^2 ) e_t^2
+ ( 2496 - 1760 \eta + 1040 \eta^2  ) e_t^4
\nonumber
\\
& \quad
+ \left[ 1920 - 768 \eta + ( 3840 - 1536 \eta ) e_t^2 \right] \sqrt{1 - e_t^2}
\Big \}
\,, 
\\
W_{\rm N} & 
= v - u + e_t \sin u
\,,
\\
W_{\rm 1PN} & 
= \frac{ 3 \xi^{2/3} }{ 1 - e_t^2 }  ( v - u + e_t \sin u )
\,,
\\
W_{\rm 2PN} & 
= \frac{ \xi ^{4/3} }{ 32 ( 1 - e_t^2 )^2 ( 1 - e_t \cos u )^3 }
\bigg (
8 
\left[ 
78 - 28 \eta 
+ ( 51 - 26 \eta) e_t^2 
- 6 ( 5 - 2 \eta ) ( 1 - e_t^2 )^{3/2} 
\right] ( v - u ) ( 1 - e_t \cos u )^3
\nonumber
\\
& \quad
+ \Big \{
624
- 284 \eta 
+ 4 \eta^2 
+ ( 408 - 88 \eta - 8 \eta^2 ) e_t^2
- ( 60 \eta - 4 \eta^2 ) e_t^4
+ \left[
- 1872
+ 792 \eta 
- 8 \eta^2 
- ( 1224 - 384 \eta 
\right.
\nonumber
\\
& \quad
\left.
- 16 \eta^2 ) e_t^2
+ ( 120 \eta - 8 \eta^2 ) e_t^4
\right] e_t \cos u 
+ \left[ 
1872 
- 732 \eta 
+ 4 \eta^2 
+ ( 1224 - 504 \eta - 8 \eta^2 ) e_t^2
- ( 60 \eta - 4 \eta^2) e_t^4 
\right] 
\nonumber
\\
& \quad
\times ( e_t \cos u )^2 
+ \left[ 
- 624 
+ 224 \eta
- ( 408 - 208 \eta ) e_t^2 
\right] ( e_t \cos u )^3 
\Big \} e_t \sin u
+ \Big \{
- ( 8 + 153 \eta - 27 \eta^2 ) e_t^2
\nonumber
\\
& \quad
+ ( 4 \eta - 12 \eta^2 ) e_t^4
+ \left[ 
8 
+ 152 \eta 
- 24 \eta^2 
+ ( 8 + 146 \eta - 6 \eta^2 ) e_t^2 
\right] e_t \cos u
+ \left[ 
- 8 
- 148 \eta 
+ 12 \eta^2 
- ( \eta - 3 \eta^2 ) e_t^2  
\right] 
\nonumber
\\
& \quad
\times ( e_t \cos u )^2
\Big \} 
e_t \sin u \sqrt{1 - e_t^2} 
\bigg )
\,, 
\\
W_{\rm 3PN} & = 
\frac{ \xi^2 }{ ( 1 - e_t^2 )^3 ( 1 - e_t \cos u )^5 }
\Bigg (
\bigg \{
\frac{285}{2}
- \frac{793}{4} \eta + \frac{123}{32} \pi^2 \eta
+ 7 \eta^2
+ \left(
\frac{879}{4}
- \frac{435}{2} \eta + \frac{123}{128} \pi^2 \eta
+ 40 \eta^2
\right) e_t^2
\nonumber
\\
& \quad
+ \left(
\frac{39}{2}
- \frac{55}{4} \eta
+ \frac{65}{8} \eta^2
\right) e_t^4
+ \bigg[
- \frac{105}{2}
+ \frac{215}{3} \eta - \frac{41}{64} \pi^2 \eta
- 5 \eta^2
+ \left(
\frac{165}{2}
- \frac{451}{6} \eta + \frac{41}{64} \pi^2 \eta
- \frac{1}{2} \eta^2
\right) e_t^2
\nonumber
\\
& \quad
+ \left( 
15
- \frac{29}{2} \eta
+ \frac{11}{2} \eta^2
\right) e_t^4
\bigg] \sqrt{1 - e_t^2} 
\bigg \} ( v - u ) ( 1 - e_t \cos u )^5
+ \bigg \{
\frac{265}{2} 
- \frac{126159}{560} \eta + \frac{287}{64} \pi^2 \eta
+ \frac{133}{8} \eta^2
+ \frac{1}{24} \eta^3
\nonumber
\\
& \quad
+ \left(
\frac{959}{4}
- \frac{118663}{840} \eta - \frac{41}{128} \pi^2 \eta
+ \frac{671}{48} \eta^2
+ \frac{7}{48} \eta^3 
\right) e_t^2
+ \left(
\frac{19}{2}
- \frac{23967}{280} \eta + \frac{41}{64} \pi^2 \eta
+ \frac{525}{16} \eta^2
- \frac{11}{16} \eta^3
\right) e_t^4
\nonumber
\\
& \quad
+ \left(
\frac{185}{8} \eta
- \frac{157}{16} \eta^2
+ \frac{37}{48} \eta^3
\right) e_t^6
- \left(
\frac{23}{48} \eta
- \frac{73}{48} \eta^2
+ \frac{13}{48} \eta^3
\right) e_t^8
+ \bigg[
- \frac{1345}{2}
+ \frac{75417}{70} \eta - \frac{697}{32} \pi^2 \eta
- 65 \eta^2
- \frac{2}{3} \eta^3
\nonumber
\\
& \quad
+ \left(
- \frac{4715}{4}
+ \frac{713597}{840} \eta + \frac{41}{128} \pi^2 \eta
- \frac{2987}{24} \eta^2
+ \frac{35}{24} \eta^3
\right) e_t^2
+ \left(
- \frac{115}{2}
+ \frac{80433}{280} \eta - \frac{41}{16} \pi^2 \eta
- \frac{419}{4} \eta^2
- \frac{3}{8} \eta^3
\right) e_t^4
\nonumber
\\
& \quad
+ \left(
- \frac{541}{8} \eta
+ \frac{173}{8} \eta^2
- \frac{23}{24} \eta^3
\right) e_t^6
+ \left(
\frac{23}{24} \eta
- \frac{73}{24} \eta^2
+ \frac{13}{24} \eta^3
\right) e_t^8
\bigg] e_t \cos u
+ \bigg[
1365
- \frac{3493027}{1680} \eta + \frac{1353}{32} \pi^2 \eta
\nonumber
\\
& \quad
+ \frac{4981}{48} \eta^2
+ \frac{71}{48} \eta^3
+ \left(
\frac{4635}{2}
- \frac{1609903}{840} \eta + \frac{123}{64} \pi^2 \eta
+ \frac{3881}{12} \eta^2
- \frac{25}{6} \eta^3
\right) e_t^2
+ \left( 
135
- \frac{12769}{35} \eta + \frac{123}{32} \pi^2 \eta
\right.
\nonumber
\\
& \quad
\left.
+ \frac{1077}{8} \eta^2
+ \frac{29}{8} \eta^3
\right) e_t^4
+ \left(
\frac{1585}{24} \eta 
- \frac{145}{12} \eta^2
- \frac{2}{3} \eta^3
\right) e_t^6
+ \left(
- \frac{23}{48} \eta
+ \frac{73}{48} \eta^2
- \frac{13}{48} \eta^3
\right) e_t^8
\bigg] ( e_t \cos u )^2
\nonumber
\\
& \quad
+ \bigg[
- 1385
+ \frac{1701499}{840} \eta - 41 \pi^2 \eta
- \frac{2065}{24} \eta^2
- \frac{9}{8} \eta^3
+ \left(
- \frac{4555}{2}
+ \frac{578789}{280} \eta - \frac{287}{64} \pi^2 \eta
- \frac{2955}{8} \eta^2
+ \frac{27}{8} \eta^3
\right) e_t^2
\nonumber
\\
& \quad
+ \left(
- 155
+ \frac{62723}{280} \eta - \frac{41}{16} \pi^2 \eta
- \frac{755}{8} \eta^2
- \frac{27}{8} \eta^3
\right) e_t^4
+ \left(
- \frac{521}{24} \eta
- \frac{35}{24} \eta^2
+ \frac{9}{8} \eta^3
\right) e_t^6
\bigg] ( e_t \cos u )^3
\nonumber
\\
& \quad
+ \bigg[
\frac{1405}{2}
- \frac{837281}{840} \eta + \frac{1271}{64} \pi^2 \eta
+ \frac{1807}{48}\eta^2
+ \frac{13}{48}\eta^3
+ \left(
\frac{4475}{4}
- \frac{150683}{140} \eta + \frac{451}{128} \pi^2 \eta
+ \frac{3143}{16} \eta^2
- \frac{13}{16} \eta^3
\right) e_t^2
\nonumber
\\
& \quad
+ \left(
\frac{175}{2}
- \frac{20887}{280} \eta + \frac{41}{64} \pi^2 \eta
+ \frac{637}{16} \eta^2
+ \frac{13}{16} \eta^3
\right) e_t^4
+ \left(
\frac{1}{6} \eta
+ \frac{83}{48} \eta^2
- \frac{13}{48} \eta^3
\right) e_t^6
\bigg] ( e_t \cos u )^4
+ \bigg[
- \frac{285}{2}
+ \frac{793}{4} \eta 
\nonumber
\\
& \quad
- \frac{123}{32} \pi^2 \eta
- 7 \eta^2
+ \left(
- \frac{879}{4}
+ \frac{435}{2} \eta - \frac{123}{128} \pi^2 \eta
- 40 \eta^2
\right) e_t^2
+ \left(
- \frac{39}{2}
+ \frac{55}{4} \eta
- \frac{65}{8} \eta^2
\right) e_t^4
\bigg] ( e_t \cos u )^5 
\bigg \} e_t \sin u
\nonumber
\\
& \quad
+ \bigg \{
15
- 6 \eta
+ \left( 
\frac{91}{4}
- \frac{412487}{3360} \eta + \frac{615}{256} \pi^2 \eta
+ \frac{2845}{64} \eta^2
- \frac{113}{64} \eta^3
\right) e_t^2
+ \left(
\frac{3}{2}
+ \frac{173297}{4480} \eta - \frac{41}{64} \pi^2 \eta
- \frac{7319}{384} \eta^2
\right.
\nonumber
\\
& \quad
\left.
+ \frac{209}{128} \eta^3
\right) e_t^4
+ \left( 
- \frac{115}{24} \eta 
+ \frac{325}{48} \eta^2
- \frac{21}{16} \eta^3
\right) e_t^6
+ \left(
\frac{1}{16}\eta
- \frac{5}{16} \eta^2
+ \frac{5}{16} \eta^3
\right) e_t^8
+ \bigg[
- \frac{275}{4}
+ \frac{49477}{420} \eta - \frac{287}{128} \pi^2 \eta
\nonumber
\\
& \quad
- \frac{395}{12} \eta^2
+ \frac{3}{4} \eta^3
+ \left(
- \frac{523}{4}
+ \frac{22553}{70} \eta - \frac{779}{128} \pi^2 \eta
- \frac{3041}{32} \eta^2
+ \frac{63}{32} \eta^3
\right) e_t^2
+ \left(
- \frac{5}{2}
- \frac{52821}{1120} \eta + \frac{41}{32} \pi^2 \eta
+ \frac{67}{12} \eta^2
\right.
\nonumber
\\
& \quad
\left.
+ \frac{35}{16} \eta^3
\right) e_t^4
+ \left(
\frac{53}{12} \eta
- \frac{121}{24} \eta^2
- \frac{3}{8} \eta^3
\right) e_t^6
\bigg] e_t \cos u
+ \bigg[
\frac{533}{4}
- \frac{75337}{280} \eta + \frac{779}{128} \pi^2 \eta
+ \frac{3269}{48} \eta^2
+ \frac{5}{16} \eta^3
\nonumber
\\
& \quad
+ \left(
\frac{1127}{4}
- \frac{204563}{560} \eta + \frac{41}{8} \pi^2 \eta
+ \frac{1615}{16} \eta^2
- \frac{21}{4} \eta^3
\right) e_t^2
+ \left(
\frac{1}{2}
- \frac{14869}{2240} \eta - \frac{41}{64} \pi^2 \eta
+ \frac{1443}{64} \eta^2
- \frac{147}{64} \eta^3
\right) e_t^4
\nonumber
\\
& \quad
+ \left(
\frac{1}{8} \eta
- \frac{23}{48} \eta^2
+ \frac{7}{16} \eta^3
\right) e_t^6
\bigg] ( e_t \cos u )^2 
+ \bigg[
- \frac{541}{4}
+ \frac{18449}{84} \eta - \frac{697}{128} \pi^2 \eta
- \frac{1073}{24} \eta^2
- \frac{9}{8} \eta^3
\nonumber
\\
& \quad
+ \left(
- \frac{1169}{4}
+ \frac{52937}{210} \eta - \frac{205}{128}\pi^2 \eta
- \frac{2291}{32} \eta^2
+ \frac{165}{32} \eta^3
\right) e_t^2
+ \left(
\frac{1}{2}
+ \frac{497}{32} \eta
- \frac{533}{48} \eta^2
+ \frac{1}{2} \eta^3
\right) e_t^4
\bigg] ( e_t \cos u )^3 
\nonumber
\\
& \quad
+ \bigg[
\frac{283}{4}
- \frac{114761}{1680} \eta + \frac{205}{128} \pi^2 \eta
+ \frac{221}{24} \eta^2
+ \frac{3}{8} \eta^3
+ \left(
\frac{297}{2}
- \frac{110277}{1120} \eta + \frac{41}{256} \pi^2 \eta
+ \frac{1439}{64} \eta^2
- \frac{87}{64} \eta^3
\right) e_t^2
\nonumber
\\
& \quad
- \left( 
\frac{5}{128} \eta
- \frac{61}{384} \eta^2
+ \frac{19}{128} \eta^3
\right) e_t^4
\bigg] ( e_t \cos u )^4
- \left[ 15 - 6 \eta + ( 30 - 12 \eta ) e_t^2 \right] ( e_t \cos u )^5
\bigg \} e_t \sin u \sqrt{1 - e_t^2}
\Bigg )
\,, 
\end{align}
\end{subequations}
and
\begin{subequations}
\begin{align}
\label{phasing_eq:26}
\dot{\phi} & 
= \dot{\phi}_{\rm N} 
+ \dot{\phi}_{\rm 1PN} 
+ \dot{\phi}_{\rm 2PN} 
+ \dot{\phi}_{\rm 3PN}
\,,
\end{align}
where
\begin{align}
\dot{\phi}_{\rm N} & = 
\frac{ n \sqrt{1 - e_t^2} }{ (1 - e_t \cos u )^2 }
\,,
\\
\dot{\phi}_{\rm 1PN} & = 
\dot{\phi}_{\rm N} \times 
\frac{ \xi^{2/3} }{ (1 - e_t^2) ( 1 - e_t \cos u) }
\left[ 3  - ( 4 - \eta ) e_t^2  + ( 1 - \eta ) e_t \cos u \right]
\,,
\\
\dot{\phi}_{\rm 2PN} & = 
\dot{\phi}_{\rm N} \times 
\frac{ \xi^{4/3} }{ 12 (1 - e_t^2)^2 ( 1 - e_t \cos u)^3 }
\biggl \{ 
144
- 48 \eta 
- ( 162 + 68 \eta - 2 \eta^2 ) e_t^2
+ ( 60 + 26 \eta - 20 \eta^2 ) e_t^4
+ ( 18 \eta + 12 \eta^2 ) e_t^6
\nonumber
\\
& \quad
+ \left[
- 216 
+ 125 \eta 
+ \eta^2 
+ ( 102 + 188 \eta + 16 \eta^2 ) e_t^2
- ( 12 + 97 \eta - \eta^2 ) e_t^4 
\right] e_t \cos u 
+ \left[
108
- 97 \eta 
- 5 \eta^2
\right. 
\nonumber
\\
& \quad
\left.
+ ( 66 - 136 \eta + 4 \eta^2 ) e_t^2
- ( 48 - 17 \eta + 17 \eta^2 ) e_t^4
\right] ( e_t \cos u )^2 
+ \left[ 
- 36
+ 2 \eta
- 8 \eta^2  
- ( 6 - 70 \eta - 14 \eta^2 ) e_t^2 
\right] ( e_t \cos u )^3 
\nonumber
\\
& \quad
+ 18 
( 1 - e_t \cos u )^2 
( 1 - 2 e_t^2 + e_t \cos u )
( 5 - 2 \eta )
\sqrt{1 - e_t^2} 
\biggr \}  
\,, 
\\
\dot{\phi}_{\rm 3PN} & = 
\dot{\phi}_{\rm N} \times 
\frac{ \xi^2 }{ ( 1 - e_t^2 )^3 (1 - e_t \cos u)^5 }
\biggl \{
75
- \frac{1447}{12} \eta + \frac{205}{64} \pi^2 \eta 
+ 2 \eta^2
+ \left(
\frac{1}{2} 
- \frac{57021}{280} \eta + \frac{205}{64} \pi^2 \eta 
+ \frac{361}{6} \eta^2
- \frac{1}{3} \eta^3 
\right) e_t^2
\nonumber
\\
& \quad
+ \left(
- 50 
+ \frac{175193}{840} \eta - \frac{41}{8} \pi^2 \eta 
- \frac{317}{6} \eta^2
+ \frac{47}{12} \eta^3
\right) e_t^4
+ \left(
18 
+ \frac{2987}{210} \eta + \frac{41}{32} \pi^2 \eta 
+ \frac{127}{6} \eta^2
- \frac{25}{4} \eta^3 
\right) e_t^6
\nonumber
\\
& \quad
+ \left( 
- \frac{863}{24} \eta 
+ \eta^2 
+ \frac{47}{12} \eta^3
\right) e_t^8
+ \left( 
\frac{9}{8} \eta 
- 3 \eta^2 
- \frac{3}{4} \eta^3 
\right) e_t^{10}
+ \left[
- 285
+ \frac{488539}{840} \eta - \frac{1025}{64} \pi^2 \eta 
- \frac{367}{24} \eta^2
+ \frac{1}{24} \eta^3
\right.
\nonumber
\\
& \quad
+ \left(
- \frac{121}{2} 
+ \frac{94097}{140} \eta - \frac{451}{64} \pi^2 \eta 
- \frac{4571}{24} \eta^2
- \frac{73}{24} \eta^3 
\right) e_t^2
+ \left(
182
- \frac{114683}{168} \eta + \frac{205}{16} \pi^2 \eta 
+ \frac{1987}{24} \eta^2
+ \frac{59}{24} \eta^3 
\right) e_t^4
\nonumber
\\
& \quad
\left.
+ \left(
- 54
+ \frac{16531}{210} \eta - \frac{41}{16} \pi^2 \eta 
 - \frac{769}{24} \eta^2
- \frac{17}{8} \eta^3 
\right) e_t^6
+ \left(
\frac{379}{12} \eta
+ \frac{25}{2} \eta^2 
+ \frac{1}{6} \eta^3 
\right) e_t^8
\right] e_t \cos u
+ \left[
411
- \frac{165061}{168} \eta 
\right.
\nonumber
\\
& \quad
+ \frac{451}{16} \pi^2 \eta 
+ \frac{121}{24} \eta^2
+ \frac{3}{8} \eta^3
+ \left(
213
- \frac{268137}{280} \eta + \frac{123}{16} \pi^2 \eta 
 + \frac{7693}{24} \eta^2
- \frac{3}{8} \eta^3
\right) e_t^2
+ \left(
- 243
+ \frac{529223}{840} \eta - \frac{369}{32} \pi^2 \eta 
\right.
\nonumber
\\
& \quad
\left.
\left.
- \frac{329}{24} \eta^2
+ \frac{45}{8} \eta^3
\right) e_t^4
+ \left(
54
- \frac{43177}{840} \eta + \frac{41}{32} \pi^2 \eta 
- \frac{227}{8} \eta^2
- \frac{7}{24} \eta^3 
\right) e_t^6
- \left( 
\frac{5}{4} \eta
- \frac{3}{2} \eta^2
+ \frac{1}{3} \eta^3
\right) e_t^8
\right]( e_t \cos u )^2 
\nonumber
\\
& \quad
+ \left[
- 273
+ \frac{288269}{420} \eta - \frac{697}{32} \pi^2 \eta 
+ \frac{189}{8} \eta^2
+ \frac{55}{24} \eta^3
- \left(
281
- \frac{79717}{84} \eta + \frac{287}{32} \pi^2 \eta 
+ \frac{7705}{24} \eta^2
+ \frac{119}{24} \eta^3
\right) e_t^2
\right.
\nonumber
\\
& \quad
\left.
+ \left(
137
- \frac{102349}{420} \eta + \frac{41}{8} \pi^2 \eta 
+ \frac{53}{24} \eta^2
- \frac{55}{24} \eta^3
\right) e_t^4
+ \left(
- 18
- \frac{347}{12} \eta
+ \frac{245}{24} \eta^2
- \frac{1}{24} \eta^3
\right) e_t^6
\right] ( e_t \cos u )^3 
\nonumber
\\
& \quad
+ \left[
78
- \frac{77011}{420} \eta + \frac{451}{64} \pi^2 \eta
- \frac{281}{24} \eta^2
- \frac{13}{8} \eta^3
+ \left(
\frac{325}{2}
- \frac{92555}{168} \eta + \frac{451}{64} \pi^2 \eta
+ \frac{3083}{24} \eta^2
+ \frac{33}{8} \eta^3
\right) e_t^2
\right.
\nonumber
\\
& \quad
\left.
+ \left(
- 23
+ \frac{5699}{105} \eta - \frac{41}{32} \pi^2 \eta
+ \frac{641}{24} \eta^2
- \frac{29}{24} \eta^3
\right) e_t^4
- \left(
\frac{11}{8} \eta 
+ \frac{23}{24} \eta^2
- \frac{29}{24} \eta^3
\right) e_t^6
\right] ( e_t \cos u )^4 
+ \left[
- 6
+ \frac{139}{8} \eta 
\right.
\nonumber
\\
& \quad
\left.
- \frac{41}{64} \pi^2 \eta
- \frac{2}{3} \eta^2
- \frac{1}{3} \eta^3
+ \left(
- \frac{69}{2}
+ \frac{13537}{140} \eta - \frac{123}{64} \pi^2 \eta
- \frac{38}{3} \eta^2
+ \frac{5}{6} \eta^3
\right) e_t^2
- \left(
3
- \frac{533}{24} \eta 
+ \frac{91}{6} \eta^2
+ \eta^3
\right) e_t^4
\right] 
\nonumber
\\
& \quad
\times ( e_t \cos u )^5
+ \frac{1}{192}
\Big \{
15840
- 16064 \eta + 123 \pi^2 \eta 
+ 960 \eta^2
- ( 38400 - 38464 \eta + 246 \pi^2 \eta + 2976 \eta^2 ) e_t^2
\nonumber
\\
& \quad
+ ( 9600 - 4416 \eta - 576 \eta^2 ) e_t^4
+ \left[
9600 
- 7680 \eta
+ 1536 \eta^2
+ ( 8640 - 21472 \eta + 246 \pi^2 \eta + 1344 \eta^2 ) e_t^2
\right.
\nonumber
\\
& \quad
\left.
+ ( 7680 - 6816 \eta + 2304 \eta^2 ) e_t^4
\right] e_t \cos u
+ \left[
- 8160
+ 12512 \eta - 123 \pi^2 \eta 
- 768 \eta^2 
- ( 4800 - 5472 \eta + 1824 \eta^2 ) e_t^2
\right] 
\nonumber
\\
& \quad
\times ( e_t \cos u )^2
\Big \} ( 1 - e_t \cos u )^3 \sqrt{1 - e_t^2}
\bigg \} 
\,. 
\end{align}
\end{subequations}

In the above equations, 
the eccentric anomaly 
$u = \mathcal{U} (l, n, e_t )$ is given by inverting
the 3PN accurate Kepler equation,
Eq.~\eqref{phasing_eq:10}, connecting $l$ and $u$.
The 3PN accurate Kepler equation 
in terms of $n$ and $e_t$ reads
\begin{align}
\label{phasing_eq:27}
l & 
= u - e_t \sin u
+ \frac{ \xi^{4/3} }{ 8 \sqrt{ 1 - e_t^2 } ( 1 - e_t \cos u ) }
\left[
( 15 \eta - \eta^2 ) e_t \sin u \sqrt{ 1 - e_t^2 }
+ 12 ( 5 - 2 \eta ) ( v - u ) ( 1 - e_t \cos u ) 
\right] 
\nonumber 
\\
& \quad
+ \frac{ \xi^2 }{ 6720 ( 1 - e_t^2 )^{3/2} ( 1 - e_t \cos u )^3 }
\bigg (
\Big \{   
67200 + 143868 \eta - 4305 \pi^2 \eta - 62160 \eta^2 - 280 \eta^3 
- ( 134400 + 139896 \eta 
\nonumber
\\
& \quad
- 8610 \pi^2 \eta - 67200 \eta^2 - 3920 \eta^3 ) e_t \cos u 
+ ( 67200 - 752 \eta - 4305 \pi^2 \eta - 15260 \eta^2 - 1820 \eta^3 )
( e_t \cos u )^2
\nonumber
\\
& \quad
+ \left[
- 148960 \eta + 45500 \eta^2 - 1540 \eta^3 
+ ( 143640 \eta - 13440 \eta^2 - 3920 \eta^3 ) e_t \cos u 
- ( 1120 \eta + 11620 \eta^2 - 1820 \eta^3 ) 
\right.
\nonumber
\\
& \quad
\left.
\times ( e_t \cos u )^2
\right] e_t^2
+ ( 3220 \eta - 10220 \eta^2 + 1820 \eta^3 ) e_t^4
\Big \} 
e_t \sin u \sqrt{ 1 - e_t^2 }
+ \left[
302400 - 461440 \eta + 4305 \pi^2 \eta + 33600 \eta^2 
\right.
\nonumber
\\
& \quad
\left.
+ ( 100800 - 97440 \eta + 36960 \eta^2 ) e_t^2
\right] 
( v - u ) ( 1 - e_t \cos u )^3
\bigg )
\,. 
\end{align}
\end{widetext}
Evidently, Eq.~\eqref{phasing_eq:27} does not have a 1PN contribution,
similar to its ADM counterpart, 
given by Eq.~(52) in Ref.~\cite{DGI}.
Note also the expected differences 
in the Kepler equation 
at higher-PN orders 
between the harmonic and the ADM gauge.

In addition to the above explicit expressions for
$r$, $\dot{r}$, $\phi$, and $\dot{\phi}$, 
we also need to evaluate
the right-hand side of Eqs.~\eqref{phasing_eq:15}.
The next subsection
contains a sketch of these computations and 
the explicit final expressions.


\subsection{3.5PN accurate reactive dynamics}

Recall that the method of Ref.~\cite{DGI}
is general and can be applied,
in principle, 
to any PN accuracy.
However, in this paper, 
we study the effects of the
2.5PN and 3.5PN contributions to radiation reaction 
on the 3PN accurate conservative motion.
Accordingly, we will   
truncate away all effects that would correspond to 
higher-PN orders.
Further,
it should be noted that 
in the case of nonspinning point masses,
it will be highly difficult to go beyond 
3.5PN accuracy 
for the oscillatory effects
associated with the two-scale decomposition, 
as given by Eq.~\eqref{phasing_eq:18}.
However, secular effects
can be computed to higher-PN orders, 
e.g., see Appendix~\ref{phasing:Appendix:PNadiabaticSec}.

Below, we provide the required inputs for
the computation of the 
3.5PN accurate evolution equations for the sets
$\{ \bar{c}_{\alpha} \}$ and $\{ \tilde{c}_{\alpha} \}$,
where 
$\alpha = n$, $e_t$, $c_l$, $c_\lambda$.
Naturally,
we require $\bm{\mathcal{A}}'$
to the 3.5PN order for this purpose. 
The appropriate 3.5PN accurate expression for $\bm{\mathcal{A}}'$ 
has to be in harmonic gauge,
as our conservative 3PN dynamics
is given in the that gauge.
The expression for the relative reactive acceleration,
to the 3.5PN order, in harmonic gauge, available in 
Ref.~\cite{Pati_Will_2002_Nissanke_Blanchet_2005_KFS_2003}, 
reads
\begin{align}
\label{phasing_eq:28}
\bm{\mathcal{A}}' & 
= {\bm{\mathcal{A}}'}_{\rm 2.5PN} 
+ {\bm{\mathcal{A}}'}_{\rm 3.5PN}
\,,
\end{align}
where 
\begin{widetext}
\begin{subequations}
\label{phasing_eq:29}
\begin{align}
\label{phasing_eq:29a}
{\bm{\mathcal{A}}'}_{\rm 2.5PN} &
= \frac{8}{15} \frac{ G^2 M^2 \eta }{ c^5 r^3 }
\left[ 
\left( 9 v^2 + 17 \frac{G M}{r} \right) \dot{r} \bm{n}
- \left( 3 v^2 + 9 \frac{G M}{r} \right) \bm{v}
\right]
\,,
\\
\label{phasing_eq:29b}
{\bm{\mathcal{A}}'}_{\rm 3.5PN} &
= \frac{2}{105} \frac{ G^2 M^2 \eta }{ c^7 r^3 }
\bigg ( 
\bigg \{ 
- ( 549 + 630 \eta ) v^4
+ ( 5985 + 630 \eta ) v^2 \dot{r}^2
- 5880 \dot{r}^4  
- \frac{G M}{r} 
\left[ ( 1038 - 2534 \eta ) v^2 \right.
\nonumber 
\\
& \quad
\left. + ( 3087 + 3948 \eta ) \dot{r}^2 \right]
- \frac{G^2 M^2}{r^2} ( 5934 + 1932 \eta )
\bigg \} \dot{r} \bm{n} 
+ \bigg \{ 
( 939 + 126 \eta ) v^4
- ( 7119 + 126 \eta ) v^2 \dot{r}^2
\nonumber 
\\
& \quad
+ 6300 \dot{r}^4 
- \frac{G M}{r} 
\left[ ( 410 + 1554 \eta ) v^2 - ( 1435 + 2968 \eta ) \dot{r}^2 \right]
+ \frac{G^2 M^2}{r^2} ( 2650 + 1092 \eta ) 
\bigg \} \bm{v} 
\bigg )
\,, 
\end{align}
\end{subequations}
\end{widetext}
where 
$v^2 = \bm{v} \cdot \bm{v} = \dot{r}^2 + r^2 \dot{\phi}^2$.
Following Ref.~\cite{DGI},
we observe that 
explicit computations of
the right-hand side of 
Eqs.~\eqref{phasing_eq:15} and \eqref{phasing_eq:17}, respectively,
require only 1PN accurate expressions for the orbital elements.
This is mainly because of the fact
that we are trying to obtain the phasing to the 3.5PN order 
and the reactive dynamics only involves 2.5PN and 3.5PN contributions.
However, this does not mean 
that the orbital dynamics is only 1PN accurate.
In all expressions where they are needed,
we have to include the appropriate PN accurate contributions.
The phasing formalism allows us
to impose the fully 1PN accurate reactive dynamics
on the 3PN accurate conservative dynamics
to provide the 3.5PN accurate phasing
[see Sec.~V in Ref.~\cite{DGI} for details].

Finally, 
the evolution equations for 
$d n / dl$, 
$d e_t / dl$, 
$d c_l / dl$, and
$d c_\lambda / dl$
in terms of $u(l,n,e_t)$, $n$, and $e_t$,
follow as
\begin{widetext}
\begin{subequations}
\label{phasing_eq:30}
\begin{align}
\label{phasing_eq:30a}
\frac{dn}{dl} & 
= - \frac{ 8  \xi ^{5/3} n \eta }{ 5 }
\bigg \{
\frac{ 6 }{ \chi^3} 
- \frac{ 32 }{\chi^4}
+ \frac{ 49 - 9 e_t^2 }{ \chi^5}
- \frac{ 35 ( 1 - e_t^2 ) }{ \chi^6 }
\bigg \}
- \frac{ \xi ^{7/3} n \eta }{ 35 } 
\bigg \{
\frac{ - 360 + 1176 \eta }{ \chi^3 }
+ \frac{ 2680 - 11704 \eta }{ \chi^4 }
\nonumber
\\
& \quad
+ \big[
- 4012 + 34356 \eta + ( 36 - 756 \eta ) e_t^2 
\big] \frac{ 1 }{ \chi^5 }
+ \big[ 
1470 - 47880 \eta - ( 350 - 17080 \eta ) e_t^2 
\big] \frac{ 1 }{ \chi^6 }
+ \big[
13510 + 31780 \eta
\nonumber
\\
& \quad
- ( 24220 + 30520 \eta ) e_t^2
+ ( 10710 - 1260 \eta ) e_t^4 
\big] \frac{ 1 }{ \chi^7 }
- \frac{ ( 27594 + 5880 \eta ) ( 1 - e_t^2 )^2 }{ \chi^8 }
+ \frac{ 11760 ( 1 - e_t^2 )^3 }{ \chi^9 }
\bigg \} 
\,,
\\
\label{phasing_eq:30b}
\frac{d e_t}{dl} &
= \frac{ 8 \xi ^{5/3} \eta ( 1 - e_t^2 ) }{ 15 e_t } 
\bigg \{
\frac{ 3 }{ \chi^3 }
- \frac{ 17 }{ \chi^4 }
+ \frac{ 49 - 9 e_t^2 }{ \chi^5}
- \frac{ 35 ( 1 - e_t^2 ) }{ \chi^6 }
\bigg \}
+ \frac{ \xi^{7/3} \eta }{ 315 e_t }
\bigg \{
\big[
- 4320 + 6636 \eta + ( 5328 - 7644 \eta ) e_t^2 
\big] \frac{ 1 }{ \chi^3 }
\nonumber
\\
& \quad
+ \big[
20430 - 36176 \eta 
- ( 25470 - 41216 \eta ) e_t^2 
\big] \frac{ 1 }{ \chi^4 }
+ \big[
- 73650 + 106120 \eta
+ ( 112692 - 116200 \eta ) e_t^2
\nonumber
\\
& \quad
- ( 39042 - 10080 \eta ) e_t^4 
\big] \frac{ 1 }{ \chi^5 }
+ \big[
102312
- 154280 \eta
- ( 201264 - 216160 \eta ) e_t^2
+ ( 98952 - 61880 \eta  ) e_t^4 
\big] \frac{ 1 }{ \chi^6 } 
\nonumber
\\
& \quad
+ \big[ 
2730
+ 95340 \eta 
+ ( 210 - 186900 \eta ) e_t^2 
- ( 8610 - 87780 \eta ) e_t^4
+ ( 5670 + 3780 \eta ) e_t^6 
\big] \frac{ 1 }{ \chi^7 } 
\nonumber
\\
& \quad
- \frac{ ( 82782 + 17640 \eta ) ( 1 - e_t^2 )^3 }{ \chi^8 }
+ \frac{ 35280 ( 1 - e_t^2 )^4 }{ \chi^9 }
\bigg \}
\,,
\\
\label{phasing_eq:30c}
\frac{d c_l}{dl} &
= \frac{ 8 \xi ^{5/3} \eta \sin u }{ 15 e_t }
\bigg \{
\frac{ 12 e_t^2}{ \chi^3 }
+ \frac{ 3 - 43 e_t^2 }{ \chi^4 }
+ \frac{ - 14 + 23 e_t^2 - 9 e_t^4}{ \chi^5 }
+ \frac{ 35 ( 1 - e_t^2 )^2 }{ \chi^6 }
\bigg \}
+ \frac{ \xi^{7/3} \eta \sin u }{ 315 e_t }
\bigg \{
\frac{ ( - 4176 + 9408 \eta ) e_t^2 }{ \chi^3 }
\nonumber
\\
& \quad
+ \big[
- 4320 
+ 6636 \eta 
+ ( 23808 - 71596 \eta ) e_t^2 
\big] \frac{ 1 }{ \chi^4 }
+ \big[ 
16110 
- 29540 \eta 
- ( 23862 - 128240 \eta ) e_t^2
\nonumber
\\
& \quad
+ ( 4392 - 6300 \eta ) e_t^4 
\big] \frac{ 1 }{ \chi^5 }
+ \big[
- 57540
+ 76580 \eta
+ ( 103320 - 159880 \eta  ) e_t^2
- ( 45780 - 83300 \eta ) e_t^4 
\big] \frac{ 1 }{ \chi^6 }
\nonumber
\\
& \quad
+ \big[ 
44772
- 77700 \eta  
- ( 59934 - 151620 \eta ) e_t^2 
- ( 14448 + 70140 \eta ) e_t^4
+ ( 29610 - 3780 \eta ) e_t^6 
\big] \frac{ 1 }{ \chi^7 }
\nonumber
\\
& \quad
+ \frac{ ( 47502 + 17640 \eta ) ( 1 - e_t^2 )^3 }{ \chi^8 } 
- \frac{ 35280 ( 1 - e_t^2 )^4 }{ \chi^9 }
\bigg \}
\,,
\\
\label{phasing_eq:30d}
\frac{d c_\lambda}{dl} & 
= - \frac{ 8 \xi^{5/3} \eta \sin u }{ 15 e_t }
\bigg \{
\left[
\frac{ 3 }{ \chi^4 }
- \frac{ 14 - 9 e_t^2  }{ \chi^5 } 
+ \frac{ 35 ( 1 - e_t^2 ) }{ \chi^6 }
\right] \sqrt{ 1 - e_t^2 } 
- \frac{ 12 e_t^2 }{ \chi^3 }
- \frac{ 3 - 43 e_t^2 }{ \chi^4 }
+ \frac{ 14 - 23 e_t^2 + 9 e_t^4 }{ \chi^5 }
- \frac{ 35 ( 1 - e_t^2 )^2}{ \chi^6 } 
\bigg \}
\nonumber
\\
& \quad
+ \frac{ 2 \xi^{7/3} \eta \sin u }{ 315 e_t ( 1 - e_t^2 ) } 
\bigg \{
\bigg [ 
\big[ 
1404 
- 3318 \eta
+ ( 360 + 3066 \eta ) e_t^2 
\big] \frac{ 1 }{ \chi^4 }
+ \big[
- 4527
+ 14770 \eta
- ( 4029 + 15232 \eta ) e_t^2
\nonumber
\\
& \quad
+ ( 576 + 882 \eta ) e_t^4 
\big] \frac{ 1 }{ \chi^5 }
+ \big[
19950
- 38290 \eta
- ( 38640 - 58520 \eta ) e_t^2
+ ( 18690 - 20230 \eta ) e_t^4 
\big] \frac{ 1 }{ \chi^6 }
\nonumber
\\
& \quad
+ \big[ 
- 22386
+ 38850 \eta
+ ( 27447 - 75810 \eta  ) e_t^2
+ ( 12264 + 35070 \eta ) e_t^4
- (  17325 - 1890 \eta ) e_t^6  
\big] \frac{ 1 }{ \chi^7 }
\nonumber
\\
& \quad
- \frac{ ( 23751 + 8820 \eta ) ( 1 - e_t^2 )^3 }{ \chi^8 }
+ \frac{ 17640 ( 1 - e_t^2 )^4 }{ \chi^9 }
\bigg ] \sqrt{1 - e_t^2}
+ \big[ 
( 2448 + 4704 \eta ) e_t^2
+ ( 2088 - 4704 \eta ) e_t^4  
\big] \frac{ 1 }{ \chi^3 }
\nonumber
\\
& \quad
+ \big[ 
- 1404 
+ 3318 \eta
- ( 4332 + 39116 \eta ) e_t^2
- ( 11904 - 35798 \eta ) e_t^4 
\big] \frac{ 1 }{ \chi^4 }
+ \big[ 
4527
- 14770 \eta
- ( 14190 - 78890 \eta ) e_t^2 
\nonumber
\\
& \quad
+ ( 11859 - 67270 \eta ) e_t^4
- ( 2196 - 3150 \eta ) e_t^6 
\big] \frac{ 1 }{ \chi^5 }
+ \big[ 
- 19950
+ 38290 \eta   
+ ( 62790 - 118230 \eta ) e_t^2
\nonumber
\\
& \quad
- ( 65730 - 121590 \eta ) e_t^4
+ ( 22890 - 41650 \eta ) e_t^6  
\big] \frac{ 1 }{ \chi^6 }
+ \big[ 
22386
- 38850 \eta 
- ( 52353 - 114660 \eta ) e_t^2
\nonumber
\\
& \quad
+ ( 22743 - 110880 \eta  ) e_t^4 
+ ( 22029 + 33180 \eta ) e_t^6
- ( 14805 - 1890 \eta ) e_t^8 
\big] \frac{ 1 }{ \chi^7 }
+ \frac{ ( 23751 + 8820 \eta ) ( 1 - e_t^2 )^4 }{ \chi^8 }
\nonumber
\\
& \quad
- \frac{ 17640 ( 1 - e_t^2 )^5 }{ \chi^9 }
\bigg \}
+ \frac{ 48 \xi^{7/3} \eta ( v - u ) }{ 5 ( 1 - e_t^2 ) }
\bigg \{
\frac{ 1 }{ \chi^3 }
- \frac{ 5 }{ \chi^4 }
\bigg \} 
\,,
\end{align}
\end{subequations}
\end{widetext}
where
$\chi \equiv 1 - e_t \cos u$ and 
$u = u ( l, n, e_t )$.
Now, we are in a position to explore the secular
and periodic variations of
$c_\alpha$ to the 3.5PN order, 
which will be done in the next two subsections.


\subsection{Secular variations}

\label{phasing:SecularSec}

First, 
let us extract the secular variations of
$c_\alpha$ from Eqs.~\eqref{phasing_eq:30}.
This is achieved by using Eqs.~\eqref{phasing_eq:19a},
which implies that the secular evolution of $c_\alpha$
can be obtained by orbital averaging
the right-hand side of Eqs.~\eqref{phasing_eq:30}, using $l$.
However,
let us note the following points before we start.
It is preferable to perform the 
orbital averaging in terms of $u$ rather than $l$,
as the right-hand sides of Eqs.~\eqref{phasing_eq:30}
are explicit functions of $u$, 
and to this accuracy we can use
$dl \simeq (1 - e_t \cos u) du$,
where we benefit from the fact that
Eq.~\eqref{phasing_eq:27} for $l$ does not have a 1PN contribution.
Moreover,
the required integration over $u$
can be easily facilitated 
by the following definite integral,
available in Ref.~\cite{WW27},
\begin{multline}
\label{phasing_eq:31}
\frac{1}{2 \pi}
\int_{0}^{2 \pi}
\frac{ du }{ (1 - e_t \cos u)^{N+1} } 
\\
= \frac{ 1 }{  ( 1 - e_t^2 )^{ (N+1)/2 }  }
P_{N} 
\bigg( \frac{ 1 }{ \sqrt {1 - e_t^2} } \bigg)
\,,
\end{multline}
where $P_N$ is the Legendre polynomial.

These statements are mainly useful to obtain the 
differential equations for $\bar{n}$ and $\bar{e}_t$, namely,
$d \bar{n}/ dl$ and $d \bar{e}_t / dl$.
Further,
it is straightforward
to express these differential equations
in terms of the original time variable $t$ rather than $l$,
by using $dl = \bar{n} dt$.
In this way, 
we obtain $d \bar{n} / dt$ and $d \bar{e}_t / dt$,
which explicitly read
\begin{widetext}
\begin{subequations}
\label{phasing_eq:32}
\begin{align}
\frac{d \bar{n}}{dt} & = 
\frac{ \xi^{5/3} n^2 \eta }{ 5 (1 - e_t^2)^{7/2} }
\Big \{ 96 + 292 e_t^2 + 37 e_t^4 \Big \}
+ \frac{ \xi^{7/3} n^2 \eta }{ 280 (1 - e_t^2)^{9/2} }
\Big \{ 
20368 - 14784 \eta 
+ ( 219880 - 159600 \eta ) e_t^2
\nonumber
\\
& \quad
+ ( 197022 - 141708 \eta ) e_t^4
+ ( 11717 - 8288 \eta ) e_t^6
\Big \}
\,,
\\
\frac{d \bar{e}_t}{dt} & = 
- \frac{ \xi^{5/3} n \eta e_t }{ 15 (1 - e_t^2)^{5/2} }
\Big \{ 304 + 121 e_t^2 \Big \}
- \frac{ \xi^{7/3} n \eta e_t }{ 2520 (1 - e_t^2)^{7/2} }
\Big \{
340968 - 228704 \eta 
+ ( 880632 - 651252 \eta ) e_t^2
\nonumber
\\
& \quad
+ ( 125361 - 93184 \eta ) e_t^4
\Big \}
\,,
\end{align}
\end{subequations}
\end{widetext}
where $n$ and $e_t$, 
on the right-hand side of these equations, 
stand for $\bar{n}$ and $\bar{e}_t$, respectively.
We have checked that our above results are in excellent agreement
with equivalent expressions,
available in Refs.~\cite{BS1989,JS1992},
computed using balance arguments involving 
local radiation damping and far-zone fluxes.
As these secular evolutions of $n$ and $e_t$, 
namely, $\bar{n}$ and $\bar{e}_t$,
are crucial for the phasing,
we additionally obtain in Appendix~\ref{phasing:Appendix:PNadiabaticSec},
for the first time,
using balance arguments,
2PN accurate expressions
for $d \bar{n} / dt$ and $d \bar{e}_t / dt$ in harmonic coordinates,
providing $\mathcal{O} (c^{-9})$ corrections. 

Now, 
let us turn our attention to the secular variation 
of $c_l$ and $c_\lambda$, namely, 
$\bar{c}_l$ and $\bar{c}_\lambda$.
The arguments, 
employed in Ref.~\cite{DGI},
to show that $\bar{G}_l = 0 = \bar{G}_\lambda$,
where $G_l = F_l / n$ and $G_\lambda = F_\lambda / n$, respectively,
at the 2.5PN order
are also extendable to the 3.5PN order.
Further,
we note that the right-hand sides of
Eqs.~\eqref{phasing_eq:30c} and \eqref{phasing_eq:30d}
are functions of the form 
$\sin u \times f(\cos u)$
and
$( v - u ) \times f(\cos u)$, 
respectively,
and hence 
they are odd under $u \rightarrow - u$.
Therefore,
their average over $dl \simeq (1 - e_t \cos u) du$
also exactly vanishes,
leading to $\bar{G}_l = 0 = \bar{G}_\lambda$ 
to the 3.5PN order.
This is also consistent
with another line of reasoning,
presented in Ref.~\cite{DGI},
that involves 
the time-odd character of 
the perturbing force $\bm{\mathcal{A}}'$,     
$\partial c_1 / \partial v^i$, and
$\partial c_2 / \partial v^j$, respectively,
appearing in Eqs.~\eqref{phasing_eq:15},
ending up with the conclusion that
$d c_l / dt$ and $d c_\lambda / dt$ are time odd.
Summarizing,
we find that there are no secular evolutions
for both $c_l$ and $c_\lambda$ 
to the 1PN order of radiation reaction:
\begin{subequations}
\label{phasing_eq:33}
\begin{align}
\label{phasing_eq:33a}
\frac{ d \bar{c}_l }{dt} & = 0 
\,; 
\quad \bar{c}_l (t) = \bar{c}_l (t_{0})
\,,
\\
\label{phasing_eq:33b}
\frac{ d \bar{c}_\lambda }{dt} & = 0 
\,; 
\quad \bar{c}_\lambda (t) = \bar{c}_\lambda (t_{0})
\,.
\end{align}
\end{subequations}


\subsection{Periodic variations}

\label{phasing:PeriodicSec}

To complete this study,
we focus now our attention
on the differential equations for 
$\tilde{n}$, 
$\tilde{e}_t$, 
$\tilde{c}_l$, and
$\tilde{c}_\lambda$,
which give orbital period oscillations to our dynamical variables
at $\mathcal{O} (c^{-5})$ and $\mathcal{O} (c^{-7})$.
First, 
let us consider the differential equations 
for $\tilde{n}$ and $\tilde{e}_t$.
They are
\begin{widetext}
\begin{subequations}
\label{phasing_eq:34}
\begin{align}
\label{phasing_eq:34a}
\frac{ d \tilde{n} }{dl} & 
= - \frac{ 8  \xi ^{5/3} n \eta }{ 5 }
\bigg \{
\frac{ 6 }{ \chi^3} 
- \frac{ 32 }{\chi^4}
+ \frac{ 49 - 9 e_t^2 }{ \chi^5}
- \frac{ 35 ( 1 - e_t^2 ) }{ \chi^6 }
\bigg \}
- \frac{ \xi^{5/3} n \eta }{ 5 (1 - e_t^2)^{7/2} }
\Big \{ 96 + 292 e_t^2 + 37 e_t^4 \Big \}
- \frac{ \xi^{7/3} n \eta }{ 35 } 
\bigg \{
\frac{ - 360 + 1176 \eta }{ \chi^3 }
\nonumber
\\
& \quad
+ \frac{ 2680 - 11704 \eta }{ \chi^4 }
+ \big[
- 4012 + 34356 \eta + ( 36 - 756 \eta ) e_t^2 
\big] \frac{ 1 }{ \chi^5 }
+ \big[
1470 - 47880 \eta - ( 350 - 17080 \eta ) e_t^2 
\big] \frac{ 1 }{ \chi^6 }
\nonumber
\\
& \quad
+ \big[ 
13510 + 31780 \eta
- ( 24220 + 30520 \eta ) e_t^2
+ ( 10710 - 1260 \eta ) e_t^4 
\big] \frac{ 1 }{ \chi^7 }
- \frac{ ( 27594 + 5880 \eta ) ( 1 - e_t^2 )^2 }{ \chi^8 }
\nonumber
\\
& \quad
+ \frac{ 11760 ( 1 - e_t^2 )^3 }{ \chi^9 }
\bigg \} 
- \frac{ \xi^{7/3} n \eta }{ 280 (1 - e_t^2)^{9/2} }
\Big \{ 
20368 - 14784 \eta 
+ ( 219880 - 159600 \eta ) e_t^2
+ ( 197022 - 141708 \eta ) e_t^4
\nonumber
\\
& \quad
+ ( 11717 - 8288 \eta ) e_t^6
\Big \}
\,,
\\
\label{phasing_eq:34b}
\frac{d \tilde{e}_t }{dl} & 
= \frac{ 8 \xi ^{5/3} \eta ( 1 - e_t^2 ) }{ 15 e_t } 
\bigg \{
\frac{ 3 }{ \chi^3 }
- \frac{ 17 }{ \chi^4 }
+ \frac{ 49 - 9 e_t^2 }{ \chi^5}
- \frac{ 35 ( 1 - e_t^2 ) }{ \chi^6 }
\bigg \}
+ \frac{ \xi^{5/3} \eta e_t }{ 15 (1 - e_t^2)^{5/2} }
\Big \{ 304 + 121 e_t^2 \Big \}
+ \frac{ \xi^{7/3} \eta }{ 315 e_t }
\bigg \{
\big[ - 4320 
\nonumber
\\
& \quad
+ 6636 \eta + ( 5328 - 7644 \eta ) e_t^2 \big] \frac{ 1 }{ \chi^3 }
+ \big[
20430 - 36176 \eta 
- ( 25470 - 41216 \eta ) e_t^2 
\big] \frac{ 1 }{ \chi^4 }
+ \big[
- 73650 + 106120 \eta
\nonumber
\\
& \quad
+ ( 112692 - 116200 \eta ) e_t^2
- ( 39042 - 10080 \eta ) e_t^4 
\big] \frac{1 }{ \chi^5 }
+ \big[ 
102312
- 154280 \eta
- ( 201264 - 216160 \eta ) e_t^2
\nonumber
\\
& \quad
+ ( 98952 - 61880 \eta  ) e_t^4 
\big] \frac{ 1 }{ \chi^6 } 
+ \big[
2730
+ 95340 \eta 
+ ( 210 - 186900 \eta ) e_t^2 
- ( 8610 - 87780 \eta  ) e_t^4
+ ( 5670 + 3780 \eta ) e_t^6 
\big] \frac{ 1 }{ \chi^7 } 
\nonumber
\\
& \quad
- \frac{ ( 82782 + 17640 \eta ) ( 1 - e_t^2 )^3 }{ \chi^8 }
+ \frac{ 35280 ( 1 - e_t^2 )^4 }{ \chi^9 }
\bigg \}
+ \frac{ \xi^{7/3} \eta e_t }{ 2520 (1 - e_t^2)^{7/2} }
\Big \{
340968 - 228704 \eta 
\nonumber
\\
& \quad
+ ( 880632 - 651252 \eta ) e_t^2
+ ( 125361 - 93184 \eta ) e_t^4
\Big \}
\,,
\end{align}
\end{subequations}
where $n$ and $e_t$, 
on the right-hand side of these equations, again 
stand for $\bar{n}$ and $\bar{e}_t$,
and we recall that the right-hand sides of Eqs.~\eqref{phasing_eq:34}
are zero-average oscillatory functions of $l$.
We have already argued that 
$\bar{G}_l = 0 = \bar{G}_\lambda$ to the 1PN reactive order.
This implies that the differential equations for 
$\tilde{c}_l$ and $\tilde{c}_\lambda$ 
are identical to those for
$c_l$ and $c_\lambda$,
as given by Eqs.~\eqref{phasing_eq:30c} and \eqref{phasing_eq:30d},
but with $n$ and $e_t$ replaced by $\bar{n}$ and $\bar{e}_t$,
respectively.
Symbolically, this reads
\begin{subequations}
\label{phasing_eq:35}
\begin{align}
\label{phasing_eq:35a}
\frac{ d \tilde{c}_l }{dl} & 
= \text{RHS of Eq.}~\eqref{phasing_eq:30c}
[ n \rightarrow \bar{n}, e_t \rightarrow \bar{e}_t ]
\,,
\\
\label{phasing_eq:35b}
\frac{ d \tilde{c}_\lambda }{dl} & 
= \text{RHS of Eq.}~\eqref{phasing_eq:30d}
[ n \rightarrow \bar{n}, e_t \rightarrow \bar{e}_t ]
\,.
\end{align}
\end{subequations}

One can analytically integrate 
Eqs.~\eqref{phasing_eq:34} and \eqref{phasing_eq:35} to get
$\tilde{n}$, 
$\tilde{e}_t$, 
$\tilde{c}_l$, and
$\tilde{c}_\lambda$
as zero-average oscillatory functions of $l$.
We find, when expressed in terms of $u$,
\begin{subequations}
\label{phasing_eq:36}
\begin{align}
\label{phasing_eq:36a}
\tilde{n} & 
= \frac{ \xi^{5/3} n \eta e_t \sin u }{ 15 ( 1 - e_t^2 )^3 }
\bigg \{
\frac{ 602 + 673 e_t^2 }{ \chi }
+ \frac{ 314 - 203 e_t^2 - 111 e_t^4 }{ \chi^2 }
+ \frac{ 98 - 124 e_t^2 - 46 e_t^4 + 72 e_t^6 }{ \chi^3 }
+ \frac{ 210 ( 1 - e_t^2 )^3 }{ \chi^4 }
\bigg \}
\nonumber
\\
& \quad
+ \frac{ \xi^{5/3} n \eta }{ 5 ( 1 - e_t^2 )^{7/2} }
\Big \{ 96 + 292 e_t^2 + 37 e_t^4 \Big \}
\bigg \{
2 \tan ^{-1} \left( \frac{ \beta_t \sin u }{ 1 - \beta_t \cos u } \right)
+ e_t \sin u 
\bigg \}
\nonumber
\\
& \quad
+ \frac{ \xi ^{7/3} n \eta e_t \sin u }{ 4200 ( 1 - e_t^2 )^4 }
\bigg \{
\big[ 
827796 - 601720 \eta 
+ ( 4322828 - 3131660 \eta ) e_t^2
+ ( 1584181 - 1132320 \eta ) e_t^4 
\big] \frac{ 1 }{ \chi }
\nonumber
\\
& \quad
+ \big[
522276 - 379960 \eta
+ ( 1024628 - 737660 \eta ) e_t^2
- ( 1371149 - 993300 \eta  ) e_t^4
- ( 175755 - 124320 \eta ) e_t^6 
\big] \frac{ 1 }{ \chi^2 }
\nonumber
\\
& \quad
+ \left[ 
391116 - 339640 \eta
- ( 296974 - 440720 \eta ) e_t^2
- ( 557800 - 66920 \eta ) e_t^4
+ ( 442058 - 97440 \eta ) e_t^6
\right.
\nonumber
\\
& \quad
\left.
+ ( 21600 - 70560 \eta ) e_t^8 
\right] \frac{ 1 }{ \chi^3 }
+ \left[ 
196476 + 155400 \eta 
- ( 401058 + 899640 \eta ) e_t^2
+ ( 24318 + 1766520 \eta ) e_t^4
\right.
\nonumber
\\
& \quad
\left.
+ ( 368634 - 1455720 \eta ) e_t^6
- ( 188370 - 433440 \eta ) e_t^8
\right] \frac{ 1 }{ \chi^4 }
+ \left[ 
170856 - 504000 \eta
- ( 668304 - 1985760 \eta ) e_t^2
\right.
\nonumber
\\
& \quad
\left.
+ ( 964656 - 2903040 \eta ) e_t^4
- ( 592704 - 1834560 \eta ) e_t^6
+ ( 110376 - 383040 \eta ) e_t^8
+ ( 15120 - 30240 \eta ) e_t^{10} 
\right] \frac{ 1 }{ \chi^5 }
\nonumber
\\
& \quad
+ \frac{ ( 115080 + 117600 \eta ) ( 1 - e_t^2 )^5 }{ \chi^6 }
- \frac{ 201600 ( 1 - e_t^2 )^6 }{ \chi^7 }
\bigg \}  
+ \frac{ \xi ^{7/3} n \eta }{ 280 ( 1 - e_t^2 )^{9/2} }
\Big \{
20368 - 14784 \eta 
+ ( 219880 - 159600 \eta ) e_t^2
\nonumber
\\
& \quad
+ ( 197022 - 141708 \eta ) e_t^4
+ ( 11717 - 8288 \eta ) e_t^6
\Big \}
\bigg \{ 
2 \tan ^{-1} \left( \frac{ \beta_t \sin u }{ 1 - \beta_t \cos u } \right)
+ e_t \sin u 
\bigg \}
\,, 
\\
\label{phasing_eq:36b}
\tilde{e}_t &
= - \frac{ \xi^{5/3} \eta \sin u }{ 45 ( 1 - e_t^2 )^2 }
\bigg \{ 
\frac{ 134 + 1069 e_t^2 + 72 e_t^4 }{ \chi } 
+ \frac{ 134  + 157 e_t^2 - 291 e_t^4 }{ \chi^2 }
+ \frac{ 98 - 124 e_t^2 - 46 e_t^4 + 72 e_t^6 }{ \chi^3 }
+ \frac{ 210 ( 1 - e_t^2 )^3 }{ \chi^4 }
\bigg \}
\nonumber
\\
& \quad
- \frac{ \xi^{5/3} \eta e_t }{ 15 ( 1 - e_t^2 )^{5/2} }
\Big \{ 304 + 121 e_t^2 \Big \}   
\bigg \{ 
2 \tan^{-1} \left( \frac{ \beta_t \sin u }{ 1 - \beta_t \cos u } \right)
+ e_t \sin u
\bigg \}
- \frac{ \xi^{7/3} \eta \sin u }{ 37800 ( 1 - e_t^2 )^3 } 
\bigg \{ 
\big[ 
78768 + 1960 \eta
\nonumber
\\
& \quad
+ ( 9997134 - 7033460 \eta ) e_t^2
+ ( 9942753 - 7434560 \eta ) e_t^4
+ ( 185760 - 131040 \eta ) e_t^6 
\big] \frac{ 1 }{ \chi }
\nonumber
\\
& \quad
+ \big[ 
78768 + 1960 \eta 
+ ( 4882614 - 3602900 \eta ) e_t^2
- ( 3266727 - 2334220 \eta  ) e_t^4
- ( 1694655 - 1266720 \eta  ) e_t^6 
\big] \frac{ 1 }{ \chi^2 }
\nonumber
\\
& \quad
+ \left[
337968 - 396200 \eta
+ ( 1267458 - 233520 \eta ) e_t^2
- ( 3136260 - 1131480 \eta ) e_t^4
+ ( 1118274 + 22400 \eta ) e_t^6
\right.
\nonumber
\\
& \quad
\left.
+ ( 412560 - 524160 \eta ) e_t^8 
\right] \frac{ 1 }{ \chi^3 }
+ \left[
- 306432 + 785400 \eta
+ ( 2380266 - 3975720 \eta ) e_t^2
- ( 5302206 - 7214760 \eta ) e_t^4
\right.
\nonumber
\\
& \quad
\left.
+ ( 4689342 - 5643960 \eta ) e_t^6
- ( 1460970 - 1619520 \eta ) e_t^8 
\right] \frac{ 1 }{ \chi^4 }
+ \left[
1419768 - 1512000 \eta
- ( 6540912 - 5957280 \eta ) e_t^2
\right.
\nonumber
\\
& \quad
\left.
+ ( 11965968 - 8709120 \eta ) e_t^4
- ( 10850112 - 5503680 \eta ) e_t^6
+ ( 4867128 - 1149120 \eta ) e_t^8
- ( 861840 + 90720 \eta ) e_t^{10} 
\right] \frac{ 1 }{ \chi^5 }
\nonumber
\\
& \quad
+ \frac{ ( 345240 + 352800 \eta ) ( 1 - e_t^2 )^5 }{ \chi^6 }
- \frac{ 604800 ( 1 - e_t^2 )^6 }{ \chi^7 }
\bigg \}
- \frac{ \xi^{7/3} \eta e_t }{ 2520 ( 1 - e_t^2 )^{7/2} } 
\Big \{
340968 - 228704 \eta
\nonumber
\\
& \quad
+ ( 880632 - 651252 \eta ) e_t^2
+ ( 125361 - 93184 \eta ) e_t^4
\Big \}
\bigg \{
2 \tan^{-1} \left( \frac{ \beta_t \sin u }{ 1 - \beta_t \cos u } \right)
+ e_t \sin u 
\bigg \}
\,, 
\\
\label{phasing_eq:36c}
\tilde{c}_l & 
= - \frac{ 2 \xi^{5/3} \eta }{ 45 e_t^2 }
\bigg \{
\frac{ 144 e_t^2 }{\chi }
+ \frac{ 18 - 258 e_t^2 }{ \chi^2 }
+ \frac{ - 56 + 92 e_t^2 - 36 e_t^4 }{ \chi^3 }
+ \frac{ 105 ( 1 - e_t^2 )^2 }{ \chi^4 }
- \frac{ 1 }{ 2 ( 1 - e_t^2 )^{3/2} } 
\left[
134 + 103 e_t^2 - 252 e_t^4
\right]
\bigg \}
\nonumber
\\
& \quad
+ \frac{ \xi^{7/3} \eta }{ 4725 e_t^2 }
\bigg \{
\frac{ ( 62640 - 141120 \eta ) e_t^2 }{ \chi }
+ \big[ 
32400 - 49770 \eta
- ( 178560 - 536970 \eta ) e_t^2 
\big] \frac{ 1 }{ \chi^2 }
+ \big[
- 80550 + 147700 \eta
\nonumber
\\
& \quad
+ ( 119310 - 641200 \eta ) e_t^2
- ( 21960 - 31500 \eta ) e_t^4 
\big] \frac{ 1 }{ \chi^3 }
+ \left[
215775 - 287175 \eta
- ( 387450 - 599550 \eta ) e_t^2 
\right.
\nonumber
\\
& \quad
\left.
+ ( 171675 - 312375 \eta ) e_t^4
\right] \frac{ 1 }{ \chi^4 }
+ \left[
- 134316 + 233100 \eta
+ ( 179802 - 454860 \eta ) e_t^2
+ ( 43344 + 210420 \eta ) e_t^4
\right.
\nonumber
\\
& \quad
\left.
- ( 88830 - 11340 \eta ) e_t^6
\right] \frac{ 1 }{ \chi^5 }
- \frac{ ( 118755 + 44100 \eta ) ( 1 - e_t^2 )^3 }{ \chi^6 }
+ \frac{ 75600 ( 1 - e_t^2 )^4 }{ \chi^7 }
+ \frac{ 1 }{ 8 ( 1 - e_t^2 )^{5/2} }
\big[ 
78768 + 1960 \eta
\nonumber
\\
& \quad
+ ( 3486804 - 2000180 \eta ) e_t^2
+ ( 2246493 + 292180 \eta ) e_t^4
- ( 335790 - 968940 \eta ) e_t^6 
\big]
\bigg \}
\,, 
\\
\label{phasing_eq:36d}
\tilde{c}_\lambda & 
= \frac{2 \xi^{5/3} \eta }{ 45 e_t^2 }
\bigg\{
\left[
\frac{ 18 }{ \chi^2 }
- \frac{ 56 - 36 e_t^2 }{ \chi^3 } 
+ \frac{ 105 (1 - e_t^2) }{ \chi^4 }
\right] \sqrt{1 - e_t^2}
- \frac{144 e_t^2}{ \chi }
- \frac{ 18 - 258 e_t^2 }{ \chi^2 }
+ \frac{ 56 - 92 e_t^2 + 36 e_t^4 }{ \chi^3 }
- \frac{ 105 (1 - e_t^2)^2 }{ \chi^4 }
\nonumber
\\
& \quad
+ \frac{ 1 }{ 2 (1 - e_t^2)^2 }
\left[
\left( 134 + 103 e_t^2 - 252 e_t^4 \right) \sqrt{ 1 - e_t^2 } 
- 134 
- 295 e_t^2 
- 36 e_t^4  
\right]
\bigg\}
- \frac{ \xi^{7/3} \eta }{ 4725 e_t^2 ( 1 - e_t^2 ) }
\bigg(
\bigg\{
\big[ 
21060 - 49770 \eta 
\nonumber
\\
& \quad
+ ( 5400 + 45990 \eta ) e_t^2  
\big] \frac{ 1 }{ \chi^2 }
+ \big[ 
- 45270
+ 147700 \eta
- ( 40290 + 152320 \eta ) e_t^2 
+ ( 5760 + 8820 \eta ) e_t^4 
\big] \frac{ 1 }{ \chi^3 }
\nonumber
\\
& \quad
+ \big[ 
149625
- 287175 \eta
- ( 289800 - 438900 \eta ) e_t^2
+ ( 140175 - 151725 \eta ) e_t^4 
\big] \frac{ 1 }{ \chi^4 }
+ \big[ 
- 134316
+ 233100 \eta
\nonumber
\\
& \quad
+ ( 164682 - 454860 \eta ) e_t^2
+ ( 73584 + 210420 \eta ) e_t^4
- ( 103950 - 11340 \eta ) e_t^6 
\big] \frac{ 1 }{ \chi^5 }
- \frac{ ( 118755 + 44100 \eta ) ( 1 - e_t^2 )^3 }{ \chi^6 }
\nonumber
\\
& \quad
+ \frac{ 75600 ( 1 - e_t^2 )^4 }{ \chi^7 }
\bigg\} \sqrt{ 1 - e_t^2 }
+ \big[ 
( 73440 + 141120 \eta ) e_t^2
+ ( 62640 - 141120 \eta ) e_t^4 
\big] \frac{ 1 }{ \chi }
+ \big[
- 21060
+ 49770 \eta
\nonumber
\\
& \quad
- ( 64980 + 586740 \eta ) e_t^2 
- ( 178560 - 536970 \eta ) e_t^4 
\big] \frac{ 1 }{ \chi^2 }
+ \big[
45270
- 147700 \eta
- ( 141900 - 788900 \eta ) e_t^2
\nonumber
\\
& \quad
+ ( 118590 - 672700 \eta ) e_t^4
- ( 21960 - 31500 \eta ) e_t^6 
\big] \frac{ 1 }{ \chi^3 }
+ \big[ 
- 149625
+ 287175 \eta
+ ( 470925 - 886725 \eta ) e_t^2
\nonumber
\\
& \quad
- ( 492975 - 911925 \eta ) e_t^4
+ ( 171675 - 312375 \eta ) e_t^6 
\big] \frac{ 1 }{ \chi^4 }
+ \big[
134316
- 233100 \eta
- ( 314118 - 687960 \eta ) e_t^2
\nonumber
\\
& \quad
+ ( 136458 - 665280 \eta ) e_t^4
+ ( 132174 + 199080 \eta ) e_t^6
- ( 88830 - 11340 \eta ) e_t^8 
\big] \frac{ 1 }{ \chi^5 }
+ \frac{ ( 118755 + 44100 \eta ) ( 1 - e_t^2 )^4 }{ \chi^6 }
\nonumber
\\
& \quad
- \frac{ 75600 ( 1 - e_t^2 )^5 }{ \chi^7 }
- \frac{ 1 }{ 8 ( 1 - e_t^2 )^2 }
\Big\{
\big[
416448
+ 1960 \eta
+ ( 3202044 - 2000180 \eta ) e_t^2
+ ( 1248573 + 292180 \eta ) e_t^4
\nonumber
\\
& \quad
- ( 335790 - 968940 \eta ) e_t^6
\big] 
\sqrt{ 1 - e_t^2 }
- 416448 
- 1960 \eta 
- ( 4223964 - 1827140 \eta ) e_t^2
- ( 3117453 - 1224020 \eta ) e_t^4
\nonumber
\\
& \quad
- ( 99810 - 69300 \eta ) e_t^6
\Big\}
\bigg) 
+ \frac{48 \xi^{7/3} \eta }{ 5 ( 1 - e_t^2 ) }
\int
( v - u )
\left(
\frac{ 1 }{ \chi^3} 
- \frac{ 5 }{ \chi^4 } 
\right)
\chi \, du    
\,,
\end{align}
\end{subequations}
\end{widetext}
where $\beta_t = (1 - \sqrt{1-e_t^2}) /e_t$.
The constant contributions to the time evolution  
of $\tilde{c}_l$ and $\tilde{c}_\lambda$,
appearing in 
Eqs.~\eqref{phasing_eq:36c} and \eqref{phasing_eq:36d}, respectively,
are required to guarantee the zero-average behaviour.
We note that 
the remaining integral in Eq.~\eqref{phasing_eq:36d}
can be numerically evaluated.

The above results
also modify the temporal evolution of 
the basic angles $l$ and $\lambda$,
entering the reactive dynamics, Eqs.~\eqref{phasing_eq:full}.
According to Ref.~\cite{DGI}, 
we see from the definitions of  
$l (t)$ and $\lambda (t)$, 
given by Eqs.~\eqref{phasing_eq:14}, 
that we can also split these angles in
secular and oscillatory pieces,
denoted by $\bar{l}$, $\bar{\lambda}$,
and $\tilde{l}$, $\tilde{\lambda}$, respectively,
as
\begin{subequations}
\label{phasing_eq:37}
\begin{align}
\label{phasing_eq:37a}
l (t) & 
= \bar{l} (t) + \tilde{l} [l;\bar{c}_a (t)]
\,,
\\
\label{phasing_eq:37b}
\lambda (t) &
= \bar{\lambda} (t) + \tilde{\lambda} [l;\bar{c}_a (t)]
\,,
\end{align}
\end{subequations}
where
\begin{subequations}
\label{phasing_eq:38}
\begin{align}
\label{phasing_eq:38a}
\bar{l} (t) &
\equiv \int_{t_0}^t \bar{n} (t) dt + \bar{c}_l (t)
\,,
\\
\label{phasing_eq:38b}
\bar{\lambda} (t) &
\equiv \int_{t_0}^t [1 + \bar{k} (t)] \bar{n} (t) dt + \bar{c}_\lambda (t)
\,.
\end{align}
\end{subequations}
We note that 
$\bar{c}_l (t) = \bar{c}_l (t_{0})$ and 
$\bar{c}_\lambda (t) = \bar{c}_\lambda (t_{0})$ are constants
[see Eqs.~\eqref{phasing_eq:33}].
The oscillatory contributions to 
$l$ and $\lambda$ are given by
\begin{subequations}
\label{phasing_eq:39}
\begin{align}
\label{phasing_eq:39a}
\tilde{l} (l; \bar{c}_a ) &
= \int \frac{ \tilde{n}(l) }{n} dl + \tilde{c}_l (l)
\,,
\\
\label{phasing_eq:39b}
\tilde{\lambda} (l; \bar{c}_a ) &
= \int 
\left[ 
\frac{ \tilde{n} }{n} + \bar{k} \frac{ \tilde{n} }{n} + \tilde{k}
\right] dl
+ \tilde{c}_\lambda (l)
\,,
\end{align}
\end{subequations}
where $\tilde{k} 
\equiv ( \partial k / \partial n) \tilde{n}
+ ( \partial k / \partial e_t ) \tilde{e}_t$
denotes the oscillatory piece in $k$.

Finally, 
to complete our study of the 
oscillatory contributions 
associated with the reactive dynamics,
we compute the integrals of Eqs.~\eqref{phasing_eq:39}
and add them to the
previous results for $\tilde{c}_l (l)$ and $\tilde{c}_\lambda (l)$,
respectively.
We find
\begin{widetext}
\begin{subequations}
\label{phasing_eq:40}
\begin{align}
\label{phasing_eq:40a}
\tilde{l} (l; \bar{c}_a ) &
= \frac{ \xi^{5/3} \eta }{ 15 ( 1 - e_t^2 )^3 }
\bigg \{ 
( 602 + 673 e_t^2 ) \chi 
+ ( 314 - 203 e_t^2 - 111 e_t^4 ) \ln \chi
- ( 602 + 673 e_t^2 ) 
+ \frac{ - 98 + 124 e_t^2 + 46 e_t^4 - 72 e_t^6 }{ \chi }
\nonumber
\\
& \quad
- \frac{ 105 ( 1 - e_t^2 )^3 }{ \chi^2 } 
\bigg \}
+ \frac{ \xi^{5/3} \eta }{ 5 ( 1 - e_t^2 )^{7/2} }
\Big \{ 96 + 292 e_t^2 + 37 e_t^4 \Big \} 
\bigg \{
\int \left[
2 \tan^{-1}
\left( \frac{ \beta_t \sin u }{ 1 - \beta_t \cos u } \right)
+ e_t \sin u
\right] \chi \, du    
\bigg \}
\nonumber
\\
& \quad
+ \frac{ \xi^{7/3} \eta }{ 4200 ( 1 - e_t^2 )^4 }
\bigg \{ 
\left[ 
827796 - 601720 \eta 
+ ( 4322828 - 3131660 \eta ) e_t^2
+ ( 1584181 - 1132320 \eta ) e_t^4
\right] \chi
\nonumber
\\
& \quad
+ \left[
522276 - 379960 \eta
+ ( 1024628 - 737660 \eta ) e_t^2
- ( 1371149 - 993300 \eta ) e_t^4
- ( 175755 - 124320 \eta ) e_t^6
\right] \ln \chi
\nonumber
\\
& \quad
- 827796 
+ 601720 \eta 
- ( 4322828 - 3131660 \eta ) e_t^2 
- ( 1584181 - 1132320 \eta ) e_t^4 
+ \big[
- 391116 + 339640 \eta
\nonumber
\\
& \quad
+ ( 296974 - 440720 \eta ) e_t^2
+ ( 557800 - 66920 \eta ) e_t^4
- ( 442058 - 97440 \eta ) e_t^6
- ( 21600 - 70560 \eta ) e_t^8 
\big] \frac{ 1 }{ \chi }
\nonumber
\\
& \quad
+ \big[
- 98238 - 77700 \eta 
+ ( 200529 + 449820 \eta ) e_t^2
- ( 12159 + 883260 \eta ) e_t^4
- ( 184317 - 727860 \eta ) e_t^6
\nonumber
\\
& \quad
+ ( 94185 - 216720 \eta ) e_t^8 
\big] \frac{ 1 }{ \chi^2 }
+ \big[
- 56952 + 168000 \eta 
+ ( 222768 - 661920 \eta ) e_t^2
- ( 321552 - 967680 \eta ) e_t^4
\nonumber
\\
& \quad
+ ( 197568 - 611520 \eta ) e_t^6
- ( 36792 - 127680 \eta ) e_t^8
- ( 5040 - 10080 \eta ) e_t^{10} 
\big] \frac{ 1 }{ \chi^3}
- \frac{ ( 28770 + 29400 \eta ) ( 1 - e_t^2 )^5 }{ \chi^4 }
\nonumber
\\
& \quad
+ \frac{ 40320 ( 1 - e_t^2 )^6 }{ \chi^5 }
\bigg \} 
+ \frac{ \xi^{7/3} \eta }{ 280 ( 1 - e_t^2 )^{9/2} }
\Big \{
20368 - 14784 \eta 
+ ( 219880 - 159600 \eta ) e_t^2
+ ( 197022 - 141708 \eta ) e_t^4
\nonumber
\\
& \quad
+ ( 11717 - 8288 \eta ) e_t^6
\Big \} 
\bigg \{
\int \left[
2 \tan^{-1}
\left( \frac{ \beta_t \sin u }{ 1 - \beta_t \cos u } \right)
+ e_t \sin u
\right] \chi \, du    
\bigg \}
+ \tilde{c}_l (l)
\,,
\\
\label{phasing_eq:40b}
\tilde{\lambda} (l; \bar{c}_a ) &
= \frac{ \xi^{5/3} \eta }{ 15 ( 1 - e_t^2 )^3 }
\bigg\{
( 602 + 673 e_t^2 ) \chi
+ ( 314 - 203 e_t^2 - 111 e_t^4 ) \ln \chi
- ( 602 + 673 e_t^2 )
+ \frac{ - 98 + 124 e_t^2 + 46 e_t^4 - 72 e_t^6 }{ \chi }
\nonumber
\\
& \quad
- \frac{ 105 ( 1 - e_t^2 )^3 }{ \chi^2 }
\bigg\}
+ \frac{ \xi^{5/3} \eta }{ 5 ( 1 - e_t^2 )^{7/2} }
\Big\{ 96 + 292 e_t^2 + 37 e_t^4 \Big\} 
\bigg\{
\int \left[
2 \tan^{-1}
\left( \frac{ \beta_t \sin u }{ 1 - \beta_t \cos u } \right)
+ e_t \sin u
\right] \chi \, du    
\bigg\} 
\nonumber
\\
& \quad
+ \frac{ \xi^{7/3} \eta }{ 4200 ( 1 - e_t^2 )^4 }
\bigg\{
\left[
1595556
- 601720 \eta
+ ( 4666388 - 3131660 \eta ) e_t^2
+ ( 1543861 - 1132320 \eta ) e_t^4
\right] \chi
\nonumber
\\
& \quad
+ \left[
886836
- 379960 \eta
+ ( 652508 - 737660 \eta ) e_t^2
- ( 1363589 - 993300 \eta ) e_t^4
- ( 175755 - 124320 \eta ) e_t^6
\right] \ln \chi
\nonumber
\\
& \quad
- 1595556 
+ 601720 \eta 
- ( 4666388 - 3131660 \eta ) e_t^2
- ( 1543861 - 1132320 \eta ) e_t^4 
+ \big[
- 473436 
+ 339640 \eta
\nonumber
\\
& \quad
+ ( 401134 - 440720 \eta ) e_t^2
+ ( 596440 - 66920 \eta ) e_t^4
- ( 502538 - 97440 \eta ) e_t^6
- ( 21600 - 70560 \eta ) e_t^8 
\big] \frac{ 1 }{ \chi }
\nonumber
\\
& \quad
+ \big[
- 186438 
- 77700 \eta
+ ( 465129 + 449820 \eta ) e_t^2
- ( 276759 + 883260 \eta ) e_t^4
- ( 96117 - 727860 \eta ) e_t^6
\nonumber
\\
& \quad
+ ( 94185 - 216720 \eta ) e_t^8 
\big] \frac{ 1 }{ \chi^2 }
+ \big[ 
- 56952 
+ 168000 \eta
+ ( 222768 - 661920 \eta ) e_t^2
- ( 321552 - 967680 \eta ) e_t^4
\nonumber
\\
& \quad
+ ( 197568 - 611520 \eta ) e_t^6
- ( 36792 - 127680 \eta ) e_t^8
- ( 5040 - 10080 \eta ) e_t^{10} 
\big] \frac{ 1 }{ \chi^3 }
- \frac{ ( 28770 + 29400 \eta ) ( 1 - e_t^2 )^5 }{ \chi^4 }
\nonumber
\\
& \quad
+ \frac{ 40320 ( 1 - e_t^2 )^6 }{ \chi^5 }
\bigg\} 
+ \frac{ \xi^{7/3} \eta }{ 280 ( 1 - e_t^2 )^{9/2} }
\Big\{
47248 - 14784 \eta
+ ( 267592 - 159600 \eta ) e_t^2
+ ( 193830 - 141708 \eta ) e_t^4
\nonumber
\\
& \quad
+ ( 11717 - 8288 \eta ) e_t^6
\Big\}  
\bigg\{
\int \left[
2 \tan ^{-1}
\left( \frac{ \beta_t \sin u }{ 1 - \beta_t \cos u } \right)
+ e_t \sin u 
\right] \chi \, du    
\bigg\} 
+ \tilde{c}_\lambda (l)
\,, 
\end{align}
\end{subequations}
\end{widetext}
where $\tilde{c}_l (l)$ and $\tilde{c}_\lambda (l)$ are given by
Eqs.~\eqref{phasing_eq:36c} and \eqref{phasing_eq:36d}, respectively,
and $\beta_t = (1 - \sqrt{ 1 - e_t^2 }) /e_t$.
Note that the contributions to $\tilde{\lambda} (l)$ 
arising from the periastron advance constant $k$ 
only appear at $\mathcal{O} (c^{-7})$.

In the next section, 
we plot the current results 
obtained in these subsections
and their influences on $h_{+}$ and $h_{\times}$.


\section{Visualization of some exemplary 3.5PN accurate results}

\label{phasing:sec_visualization}

In this section, 
we present a few samples 
of the temporal evolution of $\bar{c}_\alpha$ and $\tilde{c}_\alpha$.
These ``varying constants''
cause secular and periodic variations
in the dynamical variables 
that appear in the expressions for 
$h_{+}(t)$ and $h_{\times}(t)$,
as given by Eqs.~\eqref{phasing_eq:3}.

When we evolve orbital elements 
and gravitational waveforms, 
we have to make sure 
that the eccentric orbits we study
lie inside the area of validity of our approach.
This is guaranteed 
by terminating the orbital evolution
when the 3PN accurate version of the following constraint,
discussed in Ref.~\cite{DGI},
connecting the parameters $n$ (via $\xi = G M n / c^3$) and $e_t$,
is violated:
\begin{equation}
\label{phasing_eq:41}
\frac{ \xi }{ ( 1 - e_t^2 )^{3/2} }
= \frac{ 1 }{ j^3 }
< \frac{ 1 }{ 48^{3/2} }  
\sim 3.007 \times 10^{-3}
\,,
\end{equation}
where the dimenssionless angular-momentum variable 
is given by $j = c L / (\mu G M)$.
The above inequality,
obtainable from Eqs.~(20) and (21) in Ref.~\cite{DGI},
ensures that we are staying 
sufficiently far away from the last stable orbit (LSO).
In this way,   
the usage of Eqs.~\eqref{phasing_eq:7} and \eqref{phasing_eq:8}
for the conservative orbital dynamics
and their counterparts when the dynamics is reactive
is fully justified.
According to Eq.~\eqref{phasing_eq:41},
we will terminate the orbital evolution when $j = \sqrt{48}$.
[Recall that for a test particle around a Schwarzschild black hole,
the LSO is at $j = \sqrt{12}$.]
Going beyond the above restriction
will require the effective one body approach,
being developed by Damour and his collaborators
[see related discussions in Ref.~\cite{DGI}].

We plot in this section only dimensionless quantities like $\xi$,
in terms of dimensionless variables
by making use of the following inherent scaling.
The conversion to familiar quantities 
like orbital frequency $f$ (in hertz) 
is given by 
$f \equiv n / (2 \pi) 
= c^3 \xi / (2 \pi G M)
= 3.2312 \times 10^4 \xi (M_\odot / M)$.    
This implies that for a compact binary with the total mass
$M = M_\odot$ and $\xi = 10^{-3}$, 
the orbital frequency will be $\sim 30$ Hz.

\begin{figure}
   \centering
   \includegraphics[scale=0.36]{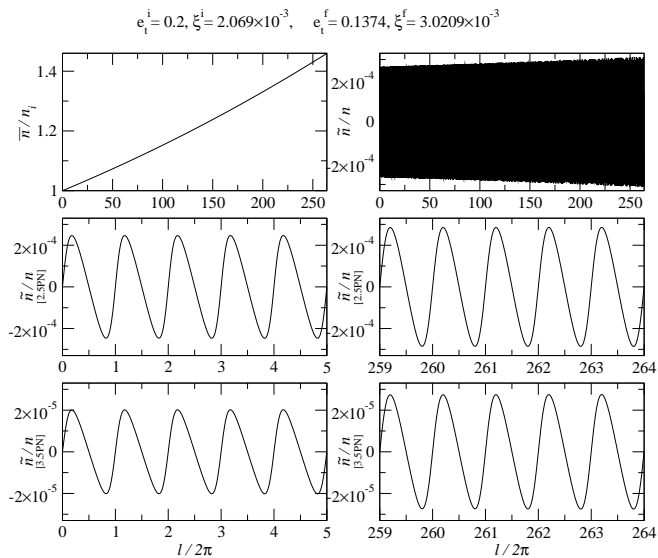}
   \caption{
   The plots for $\bar{n} / n_{i}$ and $\tilde{n} / n$
   versus $l / (2 \pi)$, which gives the number of orbital revolutions.
   The adiabatic increase of $\bar{n}$ is clearly visible in panel 1,
   and the quasi-periodic nature of the variations in $\tilde{n}$ 
   is portrayed in panels 2--6.
   These variations are governed by the reactive 
   2.5PN and 3.5PN equations of motion. 
   In the second and third row, these contributions to $\tilde{n}$ 
   are plotted individually and separated
   for the initial and final stages.
   The parameters $e_t^i$ and $e_t^f$ 
   denote initial and final values of
   the time eccentricity $e_t$, while 
   $\xi^i$ and $\xi^f$ stand for similar values of the 
   adimensional mean motion $\xi = G M n / c^3$.
   The panels are plotted for $\eta = 0.25$ 
   and the orbital evolution is terminated when $j = \sqrt{48}$.} 
  \label{phasing_fig:n_evolution}
\end{figure}

\begin{figure}
  \centering
  \includegraphics[scale=0.36]{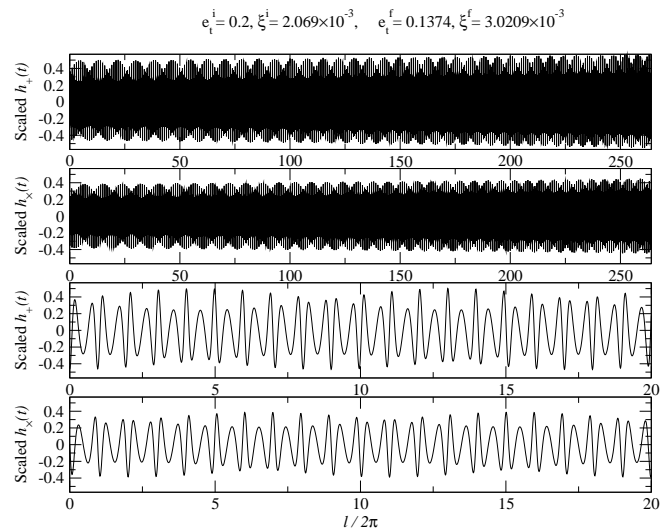}
  \caption{
  The plots for the scaled $h_{+}(t)$ and $h_{\times}(t)$
  (Newtonian in amplitude and 3.5PN in orbital motion) 
  as functions of $l / (2 \pi)$.
  The slow chirping and the amplitude modulation 
  due to the periastron precession are 
  clearly visible in the two upper panels.
  In the two bottom panels, we zoom into the initial stages 
  of the orbital evolution in order to show the effect of the 
  periodic orbital motion and the periastron advance 
  on the scaled $h_{+}(t)$ and $h_{\times}(t)$.
  The initial and final values of the relevant orbital elements
  are marked on top of the plots. 
  The panels are plotted for a binary consisting of equal masses, 
  so that $\eta = 0.25$,
  and the orbital inclination angle is given by $i = \pi / 3$.
  The orbital evolution is terminated when $j = \sqrt{48}$.} 
  \label{phasing_fig:h_evolution}
\end{figure}

In Fig.~\ref{phasing_fig:n_evolution}
we plot $\bar{n} / n_i$,
where $n_i$ is the initial value of $n$,
and $\tilde{n} / n$, 
as functions of $l / (2 \pi)$, 
which gives the evolution in terms of elapsed orbital cycles.
We clearly see  
an adiabatic increase of $\bar{n}$
as well as the quasi-periodic variations of $\tilde{n}$.
These variations are governed by the reactive 
2.5PN and 3.5PN equations of motion. 
In the second and third row of Fig.~\ref{phasing_fig:n_evolution}, 
these contributions to $\tilde{n}$ 
are plotted individually and 
separated for the initial and final stages.
These 2.5PN and 3.5PN contributions are obviously in-phase,
and we observe that the scaled 
3.5PN contributions are only by a factor of $\sim 10$ smaller
than their scaled 2.5PN counterparts.

Though,
we have all the required computations
to plot the secular and quasi-periodic variations
in $e_t$, $c_l$, and $c_\lambda$
to the 1PN reactive order,
we do not attempt it here.
As expected,
we observed the same features at the 3.5PN order,
as detailed in Figs.~2 and 3 
in Ref.~\cite{DGI} at the 2.5PN order ---
the adiabatic decrease of $\bar{e}_t$, 
the periodic variations of $\tilde{e}_t$,
no secular evolution of $\bar{c}_l$ and $\bar{c}_\lambda$, 
but periodic variations in $\tilde{c}_l$ and $\tilde{c}_\lambda$ ---
and that is the main reason for not duplicating these figures.

Finally,
we plot in Fig.~\ref{phasing_fig:h_evolution} 
the scaled $h_{+}(t)$ and $h_{\times}(t)$,
evolving under gravitational radiation reaction,
as functions of $l / (2 \pi)$.
[We factored out $G \mu / (c^4 R')$
appearing in $h_{+} |_{\rm Q}$ and $h_{\times} |_{\rm Q}$
to get the scaled waveforms.]
We employ for these figures polarization amplitudes,
which are Newtonian accurate, as given by Eqs.~\eqref{phasing_eq:3},
while the orbital motion is 3.5PN accurate.
We clearly see 
``chirping'' due to radiation damping, 
amplitude modulation due to periastron precession,
and also orbital period variations.

We note that 
Figs.~\ref{phasing_fig:n_evolution} and \ref{phasing_fig:h_evolution}
can be used to illustrate the various aspects of 
a compact binary inspiral from sources
relevant for both LIGO and LISA.
This is based on the 
above mentioned scaling argument.
Let us detail this 
in case of the following two scenarios.
For instance, 
if we choose for the total mass $M = 2.8 M_\odot$
--- a binary inspiral involving two $1.4 M_\odot$ compact objects ---
the variation of $\xi$ from 
$2.069 \times 10^{-3}$ to
$3.0209 \times 10^{-3}$           
in $\sim 264$                     
orbital cycles 
corresponds to a increasing
orbital frequency from 
$150$ Hz to $219$ Hz             
in $\sim 7.58$~s.               
Similarly, 
for the choice of $M = 10^5 M_\odot$
--- a binary inspiral involving two supermassive black holes ---
the variation of $\xi$ from 
$2.069 \times 10^{-3}$ to
$3.0209 \times 10^{-3}$         
in $\sim 264$                    
orbital cycles 
corresponds to a increasing 
orbital frequency from 
$\sim 4.2 \times 10^{-3}$ Hz to 
$\sim 6.1 \times 10^{-3}$ Hz     
in $\sim 3.1$ days.              

We conclude    
by noting that Fig.~\ref{phasing_fig:n_evolution}   
shows clearly the existence of periodic variations 
in the orbital elements --- analytically investigated 
for the first time in Ref.~\cite{DGI} 
and improved to the 3.5PN order 
in this paper.


\section{Conclusions}

\label{phasing:sec_conclusions}

Let us recapitulate.
In this paper, 
we have incorporated the 
3PN accurate conservative and the 
1PN accurate reactive dynamics
in harmonic coordinates 
into the phasing formalism to the 3.5PN order
as a natural extension of the work     
presented in Ref.~\cite{DGI}.
This extension was possible due to the very recent
determination of the 3PN accurate
generalized quasi-Keplerian parametrization
for the conservative orbital motion of
nonspinning compact binaries in eccentric orbits \cite{MGS}.
We applied the method of Ref.~\cite{DGI}
to construct, 
almost analytically,
templates for GW signals 
emitted by compact binaries 
moving in inspiralling and 
slowly precessing \emph{eccentric} orbits.
An improved 
method of variation of arbitrary constants,
explained in great detail in Ref.~\cite{DGI}, 
allowed us to combine the 
three different but relevant time scales, 
namely, those associated with the 
radial motion (orbital period),
advance of periastron,
and radiation reaction,
to the 3.5PN order
without making the usual approximation of 
treating adiabatically the radiative time scale. 
In this context, 
we recall that the two-scale decomposition 
helped us to model accurately and efficiently the time evolution
of the associated dynamical variables.
We employed harmonic coordinates
in this paper
as calculations that provided search templates for
compact binaries in quasi-circular orbits
usually employ the harmonic gauge.

The explicit computations
provided in this paper
will be required to construct 
accurate and efficient search templates
for gravitational waves from 
compact binaries of arbitrary mass ratio
moving in inspiralling eccentric orbits.
Our detailed calculations will have to be employed,
if the earth-based GW interferometers
plan to search for gravitational waves
from compact binaries with residual eccentricities,
motivated by a plethora of recent 
astrophysical investigations 
\cite{K62,MH02,Rasio2000,W03,DLK05,Page05,GZM_nature_2006,CB05,HA05_ApJ,RS_MNRAS}.
The proposed space-based GW interferometers
like LISA, BBO, and DECIGO
will have to depend on our
results to do astrophysics.
It is interesting to note
that our results will be required by LISA
to search for gravitational waves from
stellar-mass, intermediate-mass, and supermassive
black-hole binaries
as these binaries will likely to be 
in inspiralling eccentric orbits.
Another area where our computations
can be quite effective will be the
early stages of extreme mass ratio inspiral (EMRI)
as relevant for LISA.
Our current results should be also useful to benchmark
efforts that are required to obtain reliable 
EMRI templates \cite{Mino2006}.

There are many avenues 
that will require detailed investigations
in the near future
and we list only a few of them below.
In this paper,  
the conservative dynamics was restricted 
to compact binaries consisting of nonspinning point masses.
Naturally, 
it is desirable to include spin effects into our computations.
A first step in this direction 
was taken in Ref.~\cite{KG05_so_para}, 
where the 3PN accurate generalized quasi-Keplerian parametrization
for the conservative dynamics of spinning compact binaries,
moving in eccentric orbits, 
when the spin effects are restricted 
to the leading-order spin-orbit interaction, 
is presented. 
We also neglected, for simplicity,
explicit PN corrections to 
the GW polarization amplitudes of $h_{+}$ and $h_{\times}$,
and restricted them to their leading quadrupolar order.
However, it is possible
to obtain 2PN accurate corrections to these  
amplitudes,  
using Refs.~\cite{DGI,GI97}.
In order to make the numerical implementation
of our computations more efficient and accurate,
it is also desirable to provide better ways
of solving the 3PN accurate Kepler equation.
Another line of investigation should deal
with a extension of these computations
so that we have a dependable description
for the orbital evolution
near the LSO.       
Finally, these templates naturally trigger
lots of data analysis investigations
relevant for both ground-based and space-based 
GW interferometers.
Many of the above mentioned
issues are currently under investigation.


\begin{acknowledgments}

It is our pleasure to thank 
T.~Damour, B.~R.~Iyer, and G.~Sch\"afer for
illuminating discussions and persistent encouragements.
We are grateful to 
M.~Tessmer for carfully checking the typed equations.
This work is supported by the 
Deutsche Forschungsgemeinschaft (DFG) through SFB/TR7
``Gravitationswellenastronomie''.

The algebraic computations, 
appearing in this paper, 
were performed using \textsc{Maple} and \textsc{Mathematica}.

\end{acknowledgments}


\appendix

\section{Construction of an exact relation for $v - u$}

\label{phasing:Appendix:vmu}

In this appendix, 
we provide the details
involved in the derivation 
of the exact relation for $v - u$, 
which is also periodic in $u$:
\begin{align}
\label{phasing_eq:app_vmu_eq1}
v - u & = 2 \tan^{-1}
\left(
\frac{ \beta_{\phi} \sin u }{ 1 - \beta_{\phi} \cos u }
\right)
\,,
\end{align}
where $\beta_{\phi} = ( 1 - \sqrt{ 1 - e_\phi^2 } ) / e_\phi$.
This relation allows us to avoid the usage 
of the commonly employed 
infinite series expression for $v - u$, namely,
\begin{align}
\label{phasing_eq:app_vmu_eq2}
v - u & = 2 \sum_{ i = 1 }^{\infty} \frac{ \beta_\phi^i }{ i } \sin i u
\,.
\end{align}

In order to deduce Eq.~\eqref{phasing_eq:app_vmu_eq1}, 
we start from the following identity
\begin{align}
\label{phasing_eq:app_vmu_eq3}
v - u & 
\equiv 2 \tan^{-1}
\left[ \tan \left( \frac{ v - u }{ 2 } \right) \right]
\,.
\end{align}
Using therein 
\begin{align}
\label{phasing_eq:app_vmu_eq4}
\tan(\alpha - \beta) & 
= \frac{ \tan\alpha - \tan\beta }{ 1 + \tan\alpha \tan\beta }
\,,
\end{align}
leads to
\begin{align}
\label{phasing_eq:app_vmu_eq5}
v - u & 
= 2 \tan^{-1}
\left[
\frac{ \tan\frac{v}{2} - \tan\frac{u}{2} }{ 
1 + \tan\frac{v}{2} \tan\frac{u}{2} }
\right]
\,.
\end{align}
The relation connecting 
the true anomaly $v$ to 
the eccentric anomaly $u$,
as given by Eq.~\eqref{phasing_eq:9c},
is used
to replace $\tan (v/2)$ in the above equation.  
In this way, 
we obtain
\begin{align}
\label{phasing_eq:app_vmu_eq6}
v - u & 
= 2 \tan^{-1}
\left[
\frac{ ( Q_{\phi} - 1 ) \tan\frac{u}{2} }{ 
1 +  Q_{\phi} \tan^2 \frac{u}{2} }
\right]
\,,
\end{align}
where 
\begin{align}
\label{phasing_eq:app_vmu_eq7}
Q_{\phi} &
= \left( \frac{1 + e_\phi}{1 - e_\phi} \right)^{1/2} 
\,.
\end{align}
With the help of
\begin{align}
\label{phasing_eq:app_vmu_eq8}
\tan \frac{ u }{ 2 } & = \frac{ \sin u }{ 1 + \cos u }
\,,
\end{align}
we rewrite Eq.~\eqref{phasing_eq:app_vmu_eq6} as
\begin{align}
\label{phasing_eq:app_vmu_eq9}
v - u & 
= 2 \tan^{-1}
\left[
\frac{ \frac{ Q_{\phi} - 1 }{ Q_{\phi} + 1 } \sin u }{ 
1 - \frac{ Q_{\phi} - 1 }{ Q_{\phi} + 1 } \cos u }
\right]
\,.
\end{align}
Now, let us call
\begin{align}
\label{phasing_eq:app_vmu_eq10}
\beta_{\phi} & \equiv \frac{ Q_{\phi} - 1 }{ Q_{\phi} + 1 }
\,,
\end{align}
which can be simplified to  
\begin{align}
\label{phasing_eq:app_vmu_eq11}
\beta_{\phi} & 
= \frac{ 1 - \sqrt{ 1 - e_{\phi}^2 } }{ e_{\phi} }
\,.
\end{align}
Finally, 
the combination of 
Eqs.~\eqref{phasing_eq:app_vmu_eq9}--\eqref{phasing_eq:app_vmu_eq11}
directly leads to Eq.~\eqref{phasing_eq:app_vmu_eq1}.

In addition,
analog to the above derivation, 
we constructed the following exact relation, 
involving the time eccentricity $e_t$ 
instead of $e_{\phi}$, 
\begin{multline}
\label{phasing_eq:app_vmu_eq12}
2 \tan^{-1}
\left[
\left( \frac{ 1 + e_t }{ 1 - e_t } \right)^{1/2} \tan \frac{u}{2}
\right]
- u
\\
= 
2 \tan^{-1}
\left(
\frac{ \beta_{t} \sin u }{ 1 - \beta_{t} \cos u }
\right)
\,,
\end{multline}
where $\beta_{t} = ( 1 - \sqrt{ 1 - e_t^2 } ) / e_t$,
which was required 
in case of Eqs.~\eqref{phasing_eq:36a} and \eqref{phasing_eq:36b}.


\section{2PN accurate adiabatic evolution of $\bar{n}$ and $\bar{e}_t$
in harmonic coordinates}

\label{phasing:Appendix:PNadiabaticSec}

Following Ref.~\cite{DGI},
let us briefly show in this appendix
how to obtain,
in harmonic coordinates, 
the 2PN accurate secular changes in $\bar{n}$ and $\bar{e}_t$.
The PN accurate differential equations 
for $\bar{n}$ and $\bar{e}_t$ are computed 
using the heuristic arguments,
detailed in Ref.~\cite{BS1989} and in Sec.~VI in Ref.~\cite{DGI}.
In the heuristic determination
of the evolution equations for $\bar{n}$ and $\bar{e}_t$,
one employs PN accurate expressions for $n$ and $e_t$,
and the far-zone (FZ) energy and angular-momentum fluxes. 
The PN accurate expressions for 
$d \bar{n} / dt$ and $d \bar{e}_t / dt$
are then obtained by differentiating the PN accurate 
expressions for $n$ and $e_t$,
expressed in terms of $E$ and $L$,
with respect to time and then heuristically equating
the resulting time derivatives of $E$ and $L$
to the orbital averaged expressions for the
FZ energy and angular-momentum fluxes.
For the ease of implementation, 
we split the 2PN accurate computations of 
$d \bar{n} / dt$ and $d \bar{e}_t / dt$
into two parts.
The first part contains the purely 
``instantaneous'' 2PN corrections 
and the second part considers the so-called 
``tail'' contributions \cite{Def_Inst},
appearing at the 1.5PN (reactive) order
and derived for the first time in Refs.~\cite{BS1993,RS97}.
The computations to get the instantaneous contributions
begin with the 2PN corrections to the FZ fluxes,
in harmonic gauge, in terms of $r$, $\dot{r}$, and $v^2$ 
available in Ref.~\cite{GI97}.
These FZ fluxes are orbital averaged,
using the 2PN accurate 
generalized quasi-Keplerian parametrization 
for elliptical orbits in harmonic gauge, 
following the prescripton 
detailed in Ref.~\cite{BS1989}.     
We perform the orbital average by  
using an additional ingredient,
namely, the relation connecting $dl$ and $du$ to 2PN order 
in harmonic coordinates
\begin{align}
\frac{ dl }{ du } & =  
\chi      
+ \frac{ \xi^{4/3} }{ 8 \chi^2 \sqrt{ 1 - e_t^2 } } 
\bigg[
( 15 \eta - \eta^2 ) ( 1 - e_t^2 )^{3/2}
\nonumber
\\
& \quad 
+ ( 60 - 39 \eta + \eta^2 ) \chi \sqrt{ 1 - e_t^2 }
- ( 60 - 24 \eta ) 
\chi^2   
\bigg]
\,,
\end{align}
where $\chi = 1 - e_t \cos u$.
The resulting definite integrals are easily computed,
using Eq.~\eqref{phasing_eq:31}.
Now, 
we compute the time derivatives of the PN accurate expressions for 
$n$ and $e_t$ and equate the
resulting time derivatives of $E$ and $L$
to the orbital averaged
expressions for the FZ energy and angular-momentum fluxes,
respectively, to get the PN accurate expressions for
$d \bar{n} / dt$ and $d \bar{e}_t / dt$
in terms of $E$, $L$, $M$, and $\eta$. 
Finally, we use Eqs.~\eqref{phasing_eq:21}
to obtain the differential equations for 
$\bar{n}$ and $\bar{e}_t$
in terms of $\bar{n}$, $\bar{e}_t$, $M$, and $\eta$.
The resulting 2PN accurate instantaneous contributions
to $d \bar{n} / dt$ and $d \bar{e}_t / dt$,
in harmonic coordinates,
are given by
\begin{subequations}
\begin{align}
\frac{ d \bar{n} }{ dt } & = 
\bar{\xi}^{5/3} \bar{n}^2 \eta
\left \{
\dot{\bar{n}}^{\rm N}
+ \dot{\bar{n}}^{\rm 1PN}        
+ \dot{\bar{n}}^{\rm 2PN}
\right \}
\,,
\\
\frac{ d \bar{e}_t }{ dt } & =
- {\bar{\xi}}^{5/3} \bar{n} \eta \bar{e}_t
\left \{
\dot{\bar{e}}_t^{\rm N}
+ \dot{\bar{e}}_t^{\rm 1PN}      
+ \dot{\bar{e}}_t^{\rm 2PN}
\right \}
\,,
\end{align}
\end{subequations}
where the various instantaneous PN accurate corrections,
namely,
$\dot{\bar{n}}^{\rm N}$,
$\dot{\bar{n}}^{\rm 1PN}$,
$\dot{\bar{n}}^{\rm 2PN}$,
$\dot{\bar{e}}_t^{\rm N}$,
$\dot{\bar{e}}_t^{\rm 1PN}$, and
$\dot{\bar{e}}_t^{\rm 2PN}$,
read
\begin{widetext}
\begin{subequations}
\begin{align}
\dot{\bar{n}}^{\rm N} & =
\frac{1}{ 5 (1 - \bar{e}_t^2)^{7/2} }
\Bigl \{ 96 + 292 \bar{e}_t^2 + 37 \bar{e}_t^4 \Bigr \}
\,,
\\
\dot{\bar{n}}^{\rm 1PN} & =
\frac{ \bar{\xi}^{2/3} }{ 280 (1 - \bar{e}_t^2)^{9/2} }
\Bigl \{ 
20368 - 14784 \eta 
+ ( 219880 - 159600 \eta ) \bar{e}_t^2
+ ( 197022 - 141708 \eta ) \bar{e}_t^4
+ ( 11717 - 8288 \eta ) \bar{e}_t^6
\Bigr \}
\,,
\\
\dot{\bar{n}}^{\rm 2PN} & =
\frac{ \bar{\xi}^{4/3} }{ 30240 (1 - \bar{e}_t^2)^{11/2} }
\Bigl \{
12592864 - 13677408 \eta + 1903104 \eta^2
+ ( 133049696 - 185538528 \eta + 61282032 \eta^2 ) \bar{e}_t^2
\nonumber
\\
& \quad
+ ( 284496744 - 411892776 \eta + 166506060 \eta^2 ) \bar{e}_t^4
+ ( 112598442 - 142089066 \eta + 64828848 \eta^2 ) \bar{e}_t^6
\nonumber
\\
& \quad
+ ( 3523113 - 3259980 \eta + 1964256 \eta^2 ) \bar{e}_t^8
+ 3024
( 96 + 4268 \bar{e}_t^2 + 4386 \bar{e}_t^4 + 175 \bar{e}_t^6 )
( 5 - 2 \eta ) \sqrt{1 - \bar{e}_t^2}
\Bigr \}
\,,
\\
\dot{\bar{e}}_t^{\rm N} & =
\frac{ 1 }{ 15 (1 - \bar{e}_t^2)^{5/2} }
\Bigl \{ 304 + 121 \bar{e}_t^2 \Bigr \}
\,,
\\
\dot{\bar{e}}_t^{\rm 1PN} & =
\frac{ \bar{\xi}^{2/3} }{ 2520 (1 - \bar{e}_t^2)^{7/2} }
\Bigl \{
340968 - 228704 \eta 
+ ( 880632 - 651252 \eta ) \bar{e}_t^2
+ ( 125361 - 93184 \eta ) \bar{e}_t^4
\Bigr \}
\,,
\\
\dot{\bar{e}}_t^{\rm 2PN} & =
\frac{ \bar{\xi}^{4/3} }{ 30240 (1 - \bar{e}_t^2)^{9/2} }
\Bigl \{
20815216 - 25375248 \eta + 4548096 \eta^2
+ ( 87568332 - 128909916 \eta + 48711348 \eta^2 ) \bar{e}_t^2
\nonumber
\\
& \quad
+ ( 69916862 - 93522570 \eta + 42810096 \eta^2 ) \bar{e}_t^4
+ ( 3786543 - 4344852 \eta + 2758560 \eta^2 ) \bar{e}_t^6
\nonumber
\\
& \quad
+ 1008 
( 2672 + 6963 \bar{e}_t^2 + 565 \bar{e}_t^4 )
( 5 - 2 \eta ) \sqrt{1 - \bar{e}_t^2}
\Bigr \}
\,,
\end{align}
\end{subequations}
\end{widetext}
where $\bar{\xi} \equiv G M \bar{n} / c^3$.
The tail contributions to 
$d \bar{n} / dt$ and $d \bar{e}_t / dt$,
which appear at the 1.5PN order, 
are already presented in Sec.~VI in Ref.~\cite{DGI},
given by Eqs.~(70) and (71) therein.

We have checked that to the 1PN order the above contributions 
are in excellent agreement    
with Eqs.~(68) and (69) in Ref.~\cite{DGI},
which give the instantaneous 2PN accurate contributions 
to $d \bar{n} / dt$ and $d \bar{e}_t / dt$
in ADM gauge.
Note also the expected differences 
at higher-PN orders between the harmonic and the ADM gauge.
We conclude by noting that the article
providing the 3PN accurate contributions
to $d \bar{n} / dt$ and $d \bar{e}_t / dt$
is currently under preparation \cite{Bala_private}.



\begin{thebibliography}{999}
\bibitem{Def_Inst}
Following Ref.~\cite{BDIWW}, 
we term contributions to the GW amplitude
and its associated quantities
that depend only on the state of the binary at the retarded
instant as its ``instantaneous'' part, 
whereas the contributions, 
which are \emph{a priori} sensitive to 
the entire ``history'' of the binary's dynamics are termed 
as the ``tail'' contributions.
\end{thebibliography}
\end{document}